\documentclass[12pt,preprint]{aastex}

\newcommand{\giis}{GH{\sc ii}Rs}

\long\def\symbolfootnote[#1]#2{\begingroup%
\def\thefootnote{\fnsymbol{footnote}}\footnote[#1]{#2}\endgroup}
\def\kmsec{\thinspace\hbox{$\hbox{km}\thinspace\hbox{s}^{-1}$}}
\usepackage{ulem}
\slugcomment{Accepted for publication in The Astrophysical Journal}

\shorttitle{The Internal Kinematics of II Zw 40}
\shortauthors{Bordalo, Plana \& Telles}

\begin{document}


\title{The Internal Kinematics of the HII Galaxy \objectname{II Zw 40}}
\author{Vinicius Bordalo}
\affil{Observat\'orio Nacional, Rua Jos\'e Cristino, 77, Rio de Janeiro,
RJ, 20921-400, Brazil}
\email{vschmidt@on.br}

\author{Henri Plana}
\affil{Lab. Astrof\'{i}sica Te\'orica e Observacional, Univ. Estadual de
Santa Cruz, Rod. Ilheus-Itabuna km16, 45650-000 Ilh\'eus, BA, Brazil $\dagger$}
\email{plana@uesc.br}
\altaffiltext{$\dagger$}{Visiting Professor, Lab. Astrophysique de Marseille Site de Ch\^ateau-Gombert 38,  Marseille, France}
\and

\author{Eduardo Telles}
\affil{Observat\'orio Nacional, Rua Jos\'e Cristino, 77, Rio de Janeiro,
RJ, 20921-400, Brazil}
\affil{Dept. of Astronomy, University of Virginia, P.O.Box 400325,
Charlottesville, VA, 22904-4325, USA $\dagger\dagger$}
\altaffiltext{$\dagger\dagger$}{US Gemini Fellow}
\email{etelles@on.br}








\begin{abstract}

We present a study of the kinematic properties of the ionized gas
in the dominant giant H{\sc ii} region of the well known H{\sc ii}
galaxy: II Zw 40. High spatial and spectral resolution
spectroscopy has been obtained using IFU mode on the GMOS
instrument at Gemini-North\symbolfootnote[3]{Based on observations
obtained at the Gemini Observatory, which is operated by the
Association of Universities for Research in Astronomy, Inc., under
a cooperative agreement with the NSF on behalf of the Gemini
partnership: the National Science Foundation (United States), the
Science and Technology Facilities Council (United Kingdom), the
National Research Council (Canada), CONICYT (Chile), the
Australian Research Council (Australia), CNPq (Brazil) and CONICET
(Argentina)} telescope. The observations allow us to obtain the
H$\alpha$ intensity map, the radial velocity and velocity
dispersion maps as well as estimate some physical conditions in
the inner region of the starburst, such as oxygen abundance (O/H)
and electron density. We have used a set of kinematics diagnostic
diagrams, such as the intensity $versus$ velocity dispersion
(\textit{I}-$\sigma$), intensity $versus$ radial velocity
(\textit{I}-\textit{V}) and $V$-$\sigma$, for global and individual
analysis in sub-regions of the nebula. We aim to separate the main
line broadening mechanisms responsible for producing a  smooth
supersonic integrated line profile for the giant H{\sc ii} region.
Bubbles and shells driven by stellar winds and possibly
supernovae, covering a large fraction on the face of the nebula,
are identified on scales larger than 50 pc.
We found that unperturbed or ``free from shells'' regions
showing the lowest $\sigma$ values ($\sim$ 20 km s$^{-1}$)
should be good indicators for the $\sigma_{grav}$ component in
II Zw 40. The brightest central region (R $\sim$ 50 pc) is
responsible for $\sigma$ derived from a single fit to the
integrated line profile. The dominant action of gravity, and
possibly unresolved winds of young ($<$10 Myr) massive stars, in
this small region should be responsible for the characteristic
H$\alpha$ velocity profile of the starburst region as a whole
($\sigma$ = 32-35 \kmsec). Our observations show that 
the complex structure of the interstellar medium of
this galactic scale star-forming region is very similar to that of nearby
extragalactic giant H{\sc ii} regions in the Local Group galaxies.

\end{abstract}





\section{Introduction}\label{introduction}


H{\sc ii} galaxies (H{\sc ii}Gs) are dwarf galaxies undergoing a
burst of star formation. Their extensive star-forming regions,
composed of an ensemble of Super Star Clusters (SSC) and giant
H{\sc ii} regions (GH{\sc ii}Rs), dominate the optical emission.
H{\sc ii}Gs represent the simplest examples of the starburst
phenomenon occurring on galactic scales. II Zw 40 is one of the
most famous H{\sc ii}G showing an intense starburst in its central
region. Its optical spectrum is dominated by very intense emission
lines of H, He, [O {\sc ii}], [O {\sc iii}], [N {\sc ii}], [S {\sc
ii}] and [Ne {\sc iii}] superposed on a faint blue continuum,
indicating that the present rate of star formation (SF) is much
higher than the historical average
\citep{sar70,sea72,ter91,wal93,keh04}. II Zw 40 has a subsolar
oxygen abundance of 12+log(O/H)=8.07 from the data of
\cite{keh04}, typical of this class. The star formation (SF)
history of this galaxy has been the object of several studies,
ranging from the hypothesis that the star-forming episode may be
producing a first generation of stars, to a scenario in which the
history of SF consists of a continuous series of short bursts over
its lifetime. Near infrared spectra have been obtained by
\cite{van96}, who showed that II Zw 40 has low supernova rate and
that no nuclear starburst, as powerful as the present one, has
occurred in the past Giga-year. \cite{kun81} and \cite{vac92}
found Wolf-Rayet (WR) features in the optical spectrum, confirming
that the central region is very young (3-6 Myr). \cite{bec02}
found that the bright central region concentrate 75\% of the
thermal free-free emission and that the 75 pc region is formed by
four supernebula each powered by $\sim$ 600 O stars. \cite{ulv07}
also found no radio supernovae with powers greater than Cas A.

The initial trigger of the present SF episode is also a question
that remains unanswered for this class of dwarf starbursts since
most are isolated systems \citep{tt95,tm00}.  If not triggered by
external agents (i.e. interactions), star formation is caused by
internal processes manifesting in a stochastic manner
\citep{pel04}. This may be the case for the subsample of low luminosity,
compact objects, classified as type II HII galaxies by \cite{tmt97},
with no signs of morphological disturbances. However, the
more luminous HII galaxies, classified as type I, do show some signs of
extensions, fuzz, or tails in their outer envelopes, suggesting 
tidal origin. 
Over the years, only a few studies have been carried
out to determine the overall dynamics of II Zw 40. H{\sc i} velocity
mapping reveals complex structure with a subtle velocity gradient
\citep{bri88,van98}. The southeast and northwest H{\sc i} tails have
reversal rotation, suggesting that most of the material is falling
back toward the dynamical center. The best scenario for the
formation of II Zw 40 seems to be the result a collision between two
gas-rich dwarf galaxies, and the current starburst occurs in the
overlapping merging region \citep{bal82}.

In the early 70's, \citet[1972]{smi70a} first discovered broad
[OIII] and Balmer lines in GH{\sc ii}Rs. These lines have widths
which are broader (i.e. line of sight velocity dispersion $\sigma
\sim 13-25$ \kmsec) than those found in normal H{\sc ii} regions
in the Galaxy, e.g. $\sigma$ $\sim$ 5.4 km s$^{-1}$ in Orion
\citep{mun58,smi70b}. These high velocity widths imply gas motions
a few times the speed of sound. The early detailed studies
suggested that winds from WR stars should contribute to the total
kinetic energy in mass motions \citep{smi72}, and that supernova
remnants (SNRs) could play a significant role in the energetic
balance of GH{\sc ii}Rs \citep{ski84}.

In the meantime, \cite{mel77} showed a correlation between the
diameter  and the global H$\alpha$ profile widths of GH{\sc ii}Rs.
\cite{ter81} found  L(H$\beta) \propto \sigma^{\sim 4}$
which led them to propose a gravitational model for the origin of
the supersonic motions, due to the close similarities with the
parametric relations found for systems in dynamical equilibrium,
such as elliptical galaxies, globular clusters and bulges of spiral
galaxies. The first extrapolation to H{\sc ii}Gs was done by
\cite{mel88}, where a significant sample of galaxies was
observed, including II Zw 40. Recent work also confirms the
existence of the $L$-$\sigma$ relation for H{\sc ii}Gs, suggesting
that GH{\sc ii}Rs and the star-forming regions in H{\sc ii}Gs
belong to a single family of objects
\citep{roz06a,bos02,tel01,fue00}.

Several line broadening mechanisms have been proposed to interpret
the supersonic widths measured in the spectra of \giis~and H{\sc
ii}Gs. \cite{dys79} proposes that line profiles consist of two
components: one broad line component of the shocked region, and a
narrower component originated in the neighboring region which is
ionized by UV photons escaping from the broad line region. The
broad line emission is likely to arise near the shell of a hot
($10^{6}$K) highly ionized, wind-driven bubble. \cite{yor84} used
the champagne flow model \citep{ten79} -- disruption of neutral
clouds adjacent to H{\sc ii} regions-- to explain the broad lines.
The gravitational model \citep{ter81} is based on the assumption
that the dominant influence is the gravitational potential.
 In this model, the observed Gaussian line
profiles are a direct consequence of the virialized motions of the
gas plus stars complex. If one supposes a \textit{self
gravitating} system, the standard deviation of this Gaussian
profile in velocity is
\[\sigma \sim \sqrt{\frac{GM}{R}}.\]

The Cometary Stirring Model (CSM) proposed later by \cite{ten93}
improves the simpler original picture. They propose that a group
of pre-main-sequence low-mass stars of a recently formed starburst, 
move under the gravitational potential with velocity dispersion
$\sigma_{star}$.  These stars drive winds that
 stir the remaining cloud through supersonic bow,  or ``cometary'',
shocks, providing it with an average turbulent motion
$\sigma_{gas}$ $\sim$ $\sigma_{star}$. This model estimates the
total power of the shocks and associated line the luminosity which agree with the
observed relation L $\sim$ $\sigma^{4}$. Few attempts have been
made \citep[e.g.][]{ost04,cum08}, and none with a statistically
significant sample, to definitely test the hypothesis of
$\sigma_{gas} = \sigma_{star}$  which would strengthen the support
for this model.

A skeptical view of this scenario was presented by
\cite{chu94} in a detailed study of the GH{\sc ii}R 30 Doradus in
the Large Magellanic Cloud (LMC). They showed that a smooth
integrated velocity profile in the inner region (135 $\times$ 135
pc)  results from a complex velocity field, which demonstrates
the futility of using global velocity profiles to infer the
physical origin of the gas motion. The dominant contributor to the
global velocity dispersion, as proposed by the authors, at least in
the bright core, is the superposition of individual expanding
shells. Consequently, the most important physical mechanisms would
be the stellar winds from OB associations combined with
champagne-type flows and possibly supernovae.

This long disagreement in the literature has also been enlightened
by works favored by high spectral, spatial as well as
bi-dimensional spectroscopy of \giis~and the nearest H{\sc
ii}Gs. \cite{mun96} analyzed the velocity field of NGC 604 and NGC
588 in the galaxy M33. They show that, in both cases, the peaks of
high velocity dispersions are systematically concentrated in
regions of faint nebular emission. On the other hand, there is a
smooth low velocity dispersion component which permeates all
regions of the nebula. They proposed a simple criterion to
identify expanding shells in the intensity vs. $\sigma$ diagram.
This smooth low velocity dispersion component is what one measures
in the integrated spectrum of a GH{\sc ii}R. \cite{yan96} came up
with basically the same conclusions in a similar study of NGC 604.
The results found by \cite{tel01} using Echelle long-slit
spectroscopy of a small sample of H{\sc ii}Gs point in the same
direction. Despite the complex velocity fields found in such
objects, they show that the brightest knots dominate the global
luminosity and velocity dispersion. 

Though much effort has been spent to understand the kinematics of
the gas-stars complex in GH{\sc ii}Rs and H{\sc ii}Gs, a clear
picture is still absent, given the complexity of these systems. The
puzzle can be noted in the follow question: {\it Why should one find
such parametric relations in these violent environments where
massive star evolution seems to play a significant, maybe dominant,
role?} Here, we aim to bring some insight to this question by extending the
bi-dimensional spectroscopy study of Local GH{\sc ii}Rs to this
nearby prototypical H{\sc ii}G, with comparable spatial and spectral
resolution. The kinematics of the ionized gas in the very center of
II Zw 40 is investigated in some detail.

II Zw 40 has coordinates R.A. 05$^h$ 55$^m$ 42$^s$.6, DEC. +03$^o$
23$^m$ 32$\arcsec$ (J2000.0), a V = 15 mag \citep{tt97}.  It lies
at a low galactic latitude of $\sim
-10$ degrees. The observed radial velocity of 778 \kmsec~  and a
cosmological (corrected from heliocentric to cosmic microwave background frame)
recession velocity of 845 \kmsec, puts it
at a distance of 11.9 Mpc\symbolfootnote[2]{H$_{0}$ = 71
\kmsec~Mpc$^{-1}$ is throughout this paper}.

The paper is organized as follows. Section~\ref{observations}
describes the observations and reduction procedures. The results
are presented in \S~\ref{results} together with full descriptive
data analysis. In \S~\ref{discussion} we discuss our results.
Finally, we give a brief summary in \S~\ref{summary}.

\section{Observations and Data Reduction}\label{observations}

\begin{figure*}
\plotone{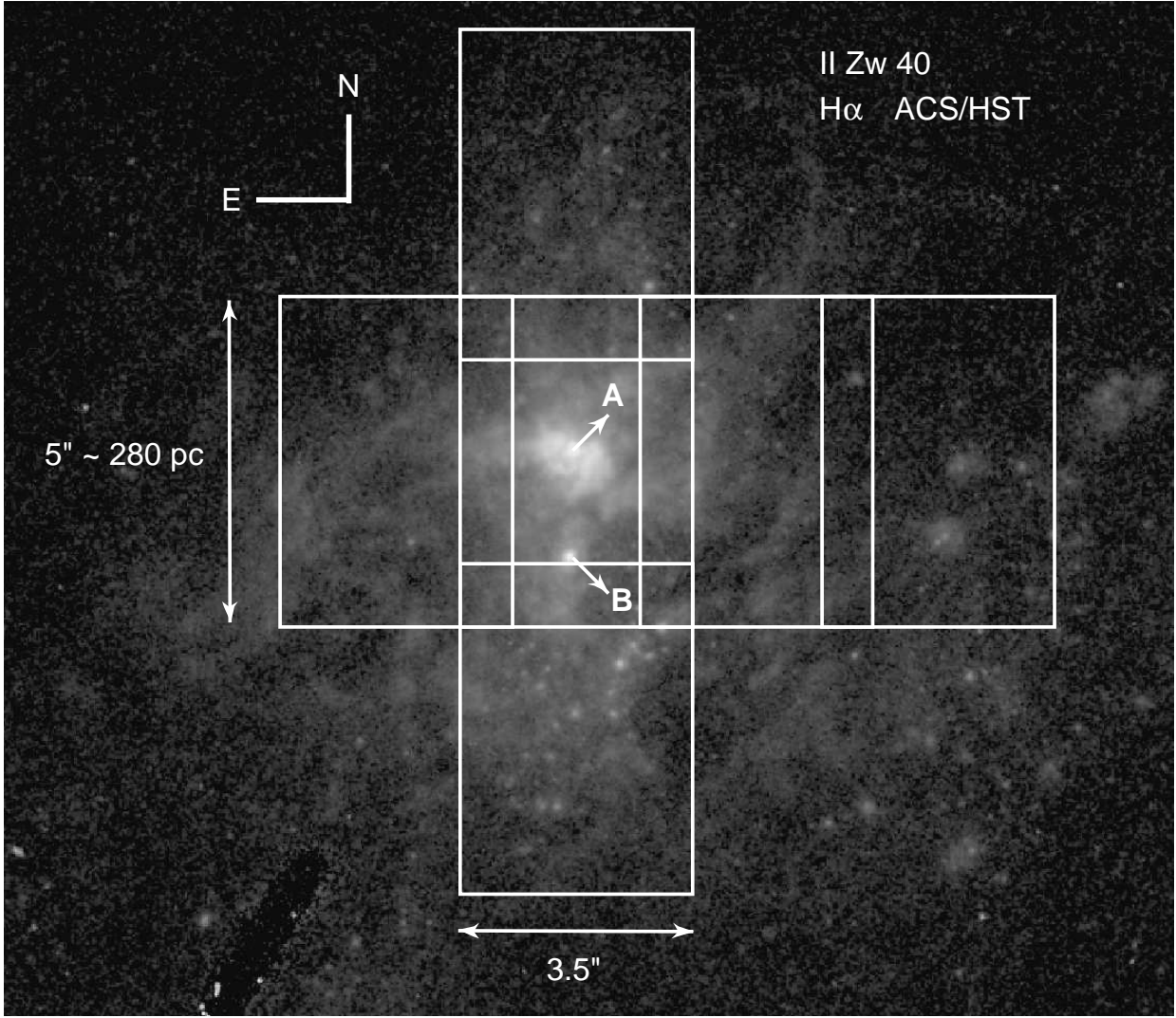} \caption{ A high resolution
H$\alpha$ image (0.027\arcsec/pixel) obtained with ACS/HST. The
GMOS/IFU fields are shown over the central region of the galaxy
(HST's archive data from the proposal 9739 by Rupali
Chandar; see also Figure 1 in \cite{van08}).\label{fig1}}
\end{figure*}

The observations of II Zw 40 were done during the run
GN-2003B-Q-26 on 2003 December 26. We used GMOS equipped with an
Integral Field Unit (IFU) on the Gemini-North Telescope. The IFU was
used in one-slit mode giving a 3.5 $\times$ 5 arcsec$^{2}$ field
of view. Each lens in its original hexagonal shape has
0.2$\arcsec$ nominal size projected in the sky. In this case, the
IFU directed to the science target has 20 $\times$ 25 lenses, in a
total of 500 lenses. A set of 249 lenses, 1 arcmin away from the
science target, provides a sky background for subtraction during
the data reduction. In order to cover a larger field, we exposed
at six different offsets overlapping the fields about 1$\arcsec$.
Figure~\ref{fig1} shows the ACS/HST high resolution H$\alpha$
image (0.027 arcsec/pixel), with
the different fields observed with GMOS/IFU shown as
rectangles over the image. The GMOS/IFU data were recorded in
three 2048 $\times$ 4608 CCDs with 13.5$\mu$m pixel size. We used
the R831 grating covering the wavelength range 5970-8080{\AA} and
providing a spectral resolution of 1.5{\AA} (FWHM) at 6580{\AA},
with corresponding $\sigma_{inst}$ = 29.0$\pm$1.0 km s$^{-1}$
($\sigma$=FWHM/2.355). The exposure time of each field was 300
seconds. The seeing remained constantly sub-arcsecond during the
observations ($\approx$ 0.5-0.6$\arcsec$).

The basic data reduction was performed using the Gemini
IRAF\symbolfootnote[3]{IRAF is distributed by the National Optical
Astronomy Observatories, which are operated by the Association of
Universities for Research in Astronomy, Inc., under cooperative
agreement with the National Science Foundation.} package developed
by the Gemini Observatory staff. A CuAr spectrum was obtained in
order to provide the wavelength calibration. We processed bias
subtraction, dome and twilight flat field correction. Each field
was calibrated in wavelength but not in flux. The final cube for
each field was re-sampled to 0.2 arcsec per pixel using GFCUBE,
resulting in a 16 $\times$ 25 $\times$ 6270 data cube. We
developed a special procedure using IRAF in order to produce the
H$\alpha$ emission, radial velocity and velocity dispersion
synthetic maps. For each data cube we used the task FITPROFS in
{\it onedspec} package to fit a single Gaussian profile to the
H$\alpha$ emission line. The task provides the H$\alpha$ central
wavelength, the FWHM and the integrated intensity in relative
counts. We run the task automatically for all spectra in the
original sampling (0.2$\arcsec$ hexagonal spaxel) and re-sampled
(0.2$\arcsec$/pixel) data cubes. There is no difference using
either data sets, nevertheless, it is more practical for the
analysis to use the re-sampled square pixel data cubes. We
constructed FITS images using H$\alpha$ line intensities, centers
and FWHM values to build our final maps. Finally, we overlapped
them to build the mosaiced field.

As an additional test, the different maps were also produced from
the original data cube using the ADHOC package \citep{bou99} in a
parallel reduction procedure in order to check the accuracy of the
data reduction. This package is designed to reduce Fabry-Perot
interferometry data and is able to handle any data cube. Both
reductions showed very similar maps, then we decided to use the
maps produced with the IRAF routines for all our further analysis.

\begin{figure}
\plotone{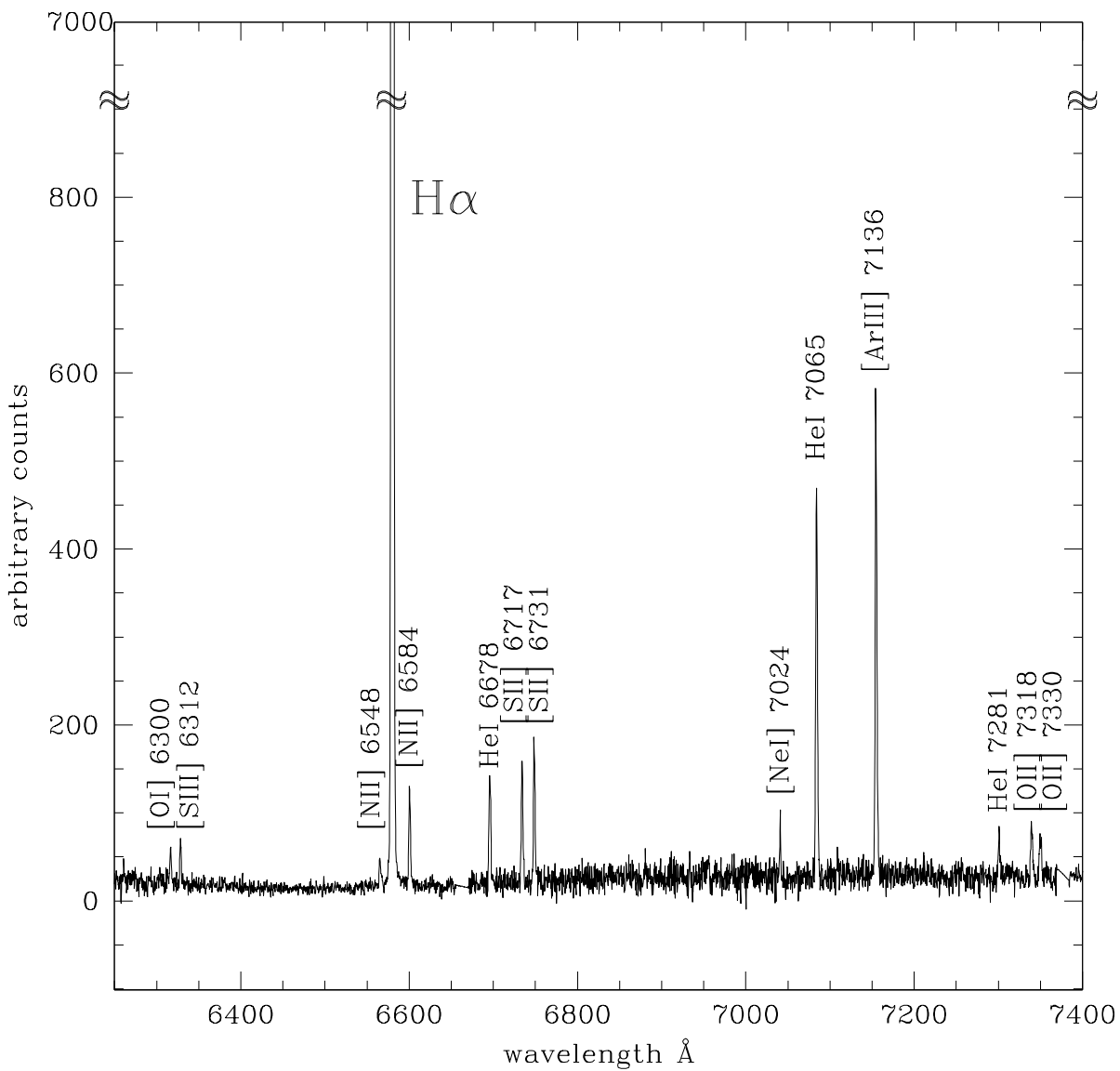} \caption{Single pixel spectrum on the bright core
centered on knot A. The labels show most of the emission lines
present on our spectrum from 6250{\AA} to 7400{\AA}. The y axis
has been cut to illustrate the relative intensity of the lines, in
particular the much larger intensity of the H$\alpha$ line which
is used in this work.} \label{spec}
\end{figure}

Figure~\ref{spec} shows a spectrum of a single pixel on the
brightest region on the main knot of II Zw 40 nebula (knot A).
This one pixel spectrum illustrates the quality of our data, and
shows that we are able to detect, resolve and measure faint lines
of [OI]$\lambda$6300, [SIII]$\lambda$6312,
[NII]$\lambda\lambda$6548,6584, HeI$\lambda$6678,
[SII]$\lambda\lambda$6717,6731, [NeI]$\lambda$7024,
HeI$\lambda$7065, [ArIII]$\lambda$7136, HeI$\lambda$7281,
[OII]$\lambda\lambda$7318, [OII]$\lambda\lambda$7330, and
[ArIII]$\lambda$7751(not shown).  The peak intensity of H$\alpha$
in this one single pixel exceeds 7000 counts and with a total flux
of 20000 counts. The S/N on the continuum around
H$\alpha$ (which is not of interest here) is about 5 on this
one pixel spectrum.  On the fainter regions, where the line widths
are still measurable, the S/N on the adjacent continuum is about
0.5, and what determines the accuracy of the line width
measurement is the peak intensity over the continuum which is
directly proportional to the total flux of the emission line. We
estimated, therefore, a conservative minimum relative flux
necessary to accurately measure a single H$\alpha$ line width of
300 counts (log F = 2.5).

The single Gaussian fit is an operational measurement procedure.
Though the major contribution comes from regions presenting line
profiles well fitted by single Gaussians, some regions of the
nebula show asymmetric line profiles or multiple components.
These measurements can map regions
of kinematic interest through line width variations, and can
reveal large scale motions through variations of the line
centroids across the nebula. When one has sufficient high spectral
resolution the analysis of a single line profile can provide
additional information constraining the larger and the smaller
scale velocity variations.

\section{Results}\label{results}

\subsection{Electron density and abundance distribution}

We briefly comment on the distribution of some derived physical
properties before we proceed with the main topic of the kinematic
properties of the starburst region in II Zw 40.  We have been able
to reliably measure and map the distribution of some weak lines in
the central field. 

\begin{figure}
\plotone{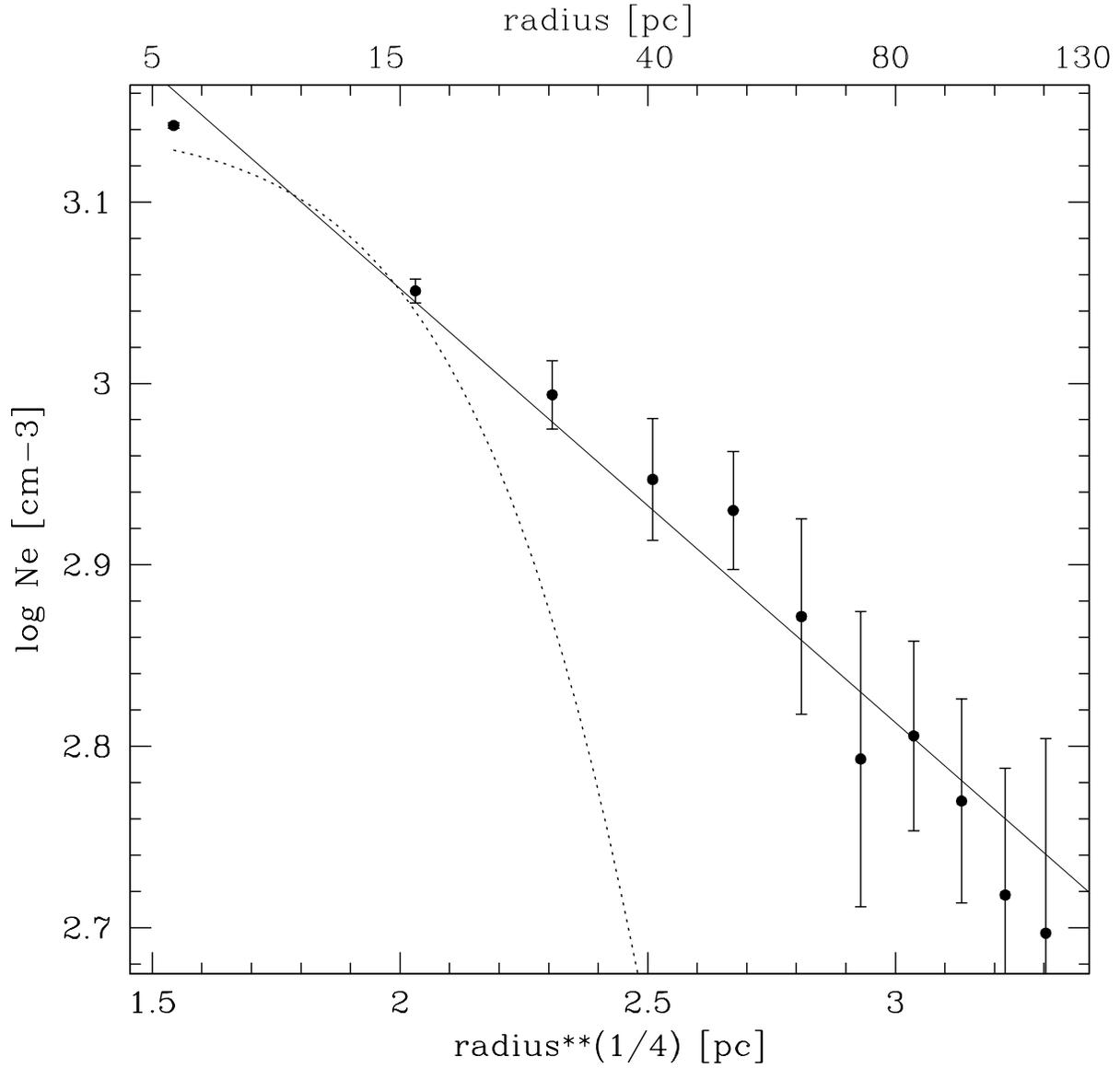} \caption{Electron density distribution in the
central field, derived from an azimuthal fit to isointensities of
the ratio of the sulfur line ([SII]$\lambda\lambda$ 6717,6731)
maps. The solid line is a r$^{1/4}$ linear fit to the data. The
dotted line is a Gaussian profile to represent the approximate
seeing during this observation.}

\label{sii}
\end{figure}

Figure~\ref{sii} shows the electron density (n$_e$) distribution
derived from  the ratio of
[SII]$\lambda\lambda$6717,6731 lines \citep{dop03}. This plot was
produced by azimuthally averaging the 2D map centered on the peak
distribution of the sulfur line ratio which is coincident with the
peak H$\alpha$ line emission from knot A. Firstly, we show that
n$_e$ is not constant and ranges from $\sim$1400 cm$^{-3}$ in the
center to  $\sim$ 500 cm$^{-3}$ about 100 pc from the center of
knot A, and higher than the low density limit of 100
cm$^{-3}$.  It is also
noticeable the good fit to a r$^{1/4}$ law (solid line) to the
outer points (avoiding the dominant effect of seeing in the center
represented by the dotted line), which is usually representative
of the surface brightness distribution of stellar systems in
dynamical equilibrium.

\begin{figure}
\epsscale{0.49} \plotone{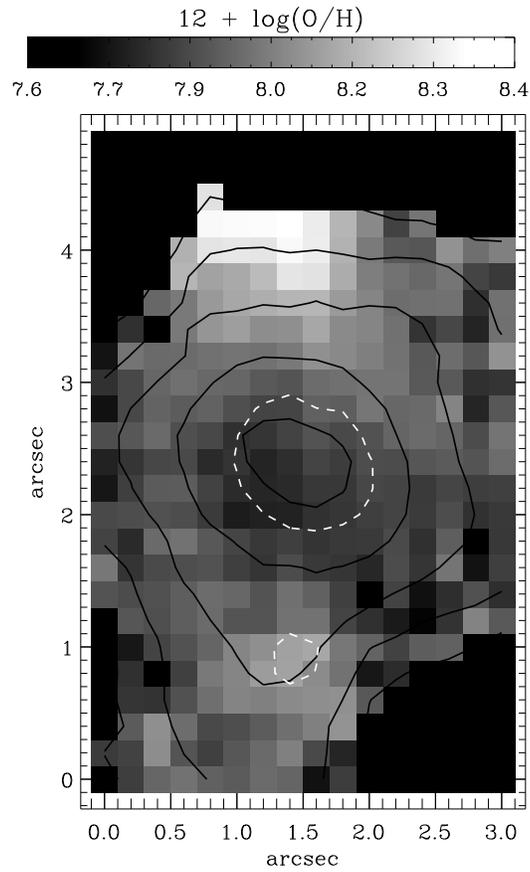} \caption{Oxygen abundance
distribution in the central field, derived from the ratio of
[NII]$\lambda$6584/H$\alpha$ \citep{den02}. The H$\alpha$ emission
contours at 68, 20, 9, 5 and 3.4\% of the peak intensity (solid
lines) and the continuum emission contours (dashed lines) at 47\%
of its peak intensity are superposed on the image.
We use  a 3$\times$3 box smooth the pixel-to-pixel variation
due to poisson noise}

\label{o_h}
\end{figure}

Figure~\ref{o_h} shows the oxygen abundance distribution in this
central field. H$\alpha$ and continuum emission are also plotted
as contours over the image. All pixels with H$\alpha$ emission
less than 3.4\% of the peak emission were erased. The O/H
abundance was derived by applying the N2 calibrator from
\cite{den02}. This is an empirical relation between the
the line ratio of [NII]$\lambda$6584/H$\alpha$ with the
oxygen abundance. Simple closed box chemical evolution models
considering the primary and secondary origin of nitrogen
predict this linear correlation between nitrogen and oxygen,
though dependent on the ionization parameter. Here, we simply
apply the empirical calibration of this relation.
We found a lower abundance over knot A, 12+log(O/H)
= 7.85, a slight increase over knot B, 12+log(O/H) = 8.0
and a significant increase in a region at NE of the knot A,
12+log(O/H) = 8.3. The statistical error of the N2 estimator
is $\delta_{(\rm{O/H})} \sim 0.15$, however relative differences
between the regions should be real.

We will not explore the physical conditions analysis here any
further. However, we may say that we have not found any relation
between the velocity dispersion and the oxygen abundance
or n$_e$.

\subsection{The H$\alpha$ Integrated Line Profile}

We now compare the integrated line width for the main starburst
region in II Zw 40 derived here, with the inner core line width
obtained with the high dispersion fiber spectrograph FEROS
\citep{kau99}, on the 1.52m telescope (ESO)\symbolfootnote[4]{All the
FEROS data and analysis were presented in V. Bordalo's Ms. Sc.
dissertation (2004) directed by Dr. E. Telles at Observat\'orio
Nacional - MCT/Brazil, and will be published in a forthcoming
paper.}. We combined 181 spectra from our GMOS data
cube, around the inner core in a synthetic aperture of a diameter
of 2.7$\arcsec$ (equivalent to 150 pc) which corresponds to the
FEROS fiber diameter.

Figure~\ref{feros} shows the line profiles, their single Gaussian
fits and the velocity dispersions derived from the data of these
two very different instruments. Our simulated spectrum from
GMOS/IFU data shows an observed FWHM = 2.34{\AA}, while the FEROS
spectrum shows FWHM = 1.83{\AA}. After the respective instrumental
broadening, and the thermal broadening corrections (for $T_{e}$ =
10$^{4}$K, FWHM$_{th}$ = 0.47{\AA} or $\sigma_{th}$ = 9.1 \kmsec),
the velocity dispersions derived were virtually the same, $\sigma$ = 34 \kmsec.

II Zw 40 presents an integrated supersonic line width  with a
derived high velocity dispersion reaching 34 \kmsec, which is
significantly higher than 30 Dor with $\sigma =$26 \kmsec, the
most powerful GH{\sc ii}R of our Local Group, \citep{chu94,mel99}.

\subsection{The H$\alpha$ Flux Monochromatic Map}

The H$\alpha$ flux monochromatic map is a result of the emission line
intensity measurements across the observed field. We fitted single
Gaussians to the H$\alpha$ line profiles to derive their
respective total fluxes, using the procedure discussed in
Section~\ref{observations}. The H$\alpha$ monochromatic 
(total flux from the single Gaussian fit)
contour map is presented superposed on the velocity dispersion and radial
velocity maps (Figures~\ref{disp} and~\ref{vel}) in six contours
levels (68, 20, 9, 5, 3.4 and 2.6\% of the peak intensity). The
overall H$\alpha$ distribution in the synthetic map obtained from
GMOS/IFU agrees very well with the narrow band H$\alpha$ image
from ACS/HST at higher resolution (Figure~\ref{fig1}). In the
HST's image it is possible to clearly resolve the two inner knots
(A and B), the cavities formed by low H$\alpha$ emission
maybe associated with superbubbles, and also some filaments. A
synthetic red continuum image from our IFU data, shown here only
in two contours in Figure~\ref{o_h}, also reveals the two inner
knots, separated by projected distance of 1.6$\arcsec$ ($\sim$ 90
pc). However, the Northern knot (A) is responsible for virtually
all the ionizing luminosity in the core of II Zw 40 \citep{van08}.

The starburst region revealed in II Zw 40, from the morphological
point of view, is also very similar to the largest and extensively
studied counterparts in the Local Group, such as 30 Dor, NGC 588
and NGC 604, showing complex kinematic features, along with
filamentary structure, shells, and cavities, suggesting a
common physical cause \citep{ten06}. The similarities
of the properties of the ISM in GH{\sc ii}Rs and H{\sc ii}Gs were
first investigated through empirical relations of integrated
physical parameters \citep{mel87,mel88}, particularly the fact
that both exhibit supersonic motions of the warm interstellar
medium. A more detailed analysis of spatially resolved properties
of these structures will help us understand the interplay between
the massive cluster formation and evolution with its surrounding
interstellar medium on scales of tens to hundreds of parsec.

\subsection{The H$\alpha$ Velocity Dispersion and Radial Velocity Maps}

The velocity dispersion  map is presented in
Figure~\ref{disp}, with H$\alpha$ intensity contours 
superposed. The region corresponding to the brightest
knot (inner 50 pc) presents a low to intermediate $\sigma$ value
(25-35 \kmsec). There are, at least, three well defined high
dispersion (45-65 km s$^{-1}$) regions  $\sim$100 pc
from the brightest knot. An almost constant velocity
dispersion field ($\leq$ 35 \kmsec), which contains most H$\alpha$
contour levels, is present in a region with 100$\times$100
pc$^{2}$ at NW of the brightest knot. At least two regions of low
velocity dispersion (15-25 \kmsec) are also found in the field,
strikingly in between regions of high dispersion. A third low
dispersion region is seen at extreme West, 300 pc far from the core,
but we preferred to discard this from the analysis due to its low
S/N.

The observed H$\alpha$ radial velocity map is shown in
Figure~\ref{vel}. There is no sign of an overall rotation in the
main starburst region of II Zw 40 (~500$\times$500 pc$^{2}$).
 On the other hand, discrete
regions can reach high values of radial velocity,
suggesting local expanding motions, although 
the total velocity range does not exceed 60 \kmsec.

\subsection{$Intensity$-$\sigma$ and $Intensity$-$Velocity$
Diagrams}

An efficient method to identify kinematic features in GH{\sc ii}Rs, detect the
massive clusters, their effect on the ISM, and their star formation rates, is by
the analysis of the $Intensity$-$\sigma$ ($I$-$\sigma$) and
$Intensity$-$Velocity$ ($I$-$V$) diagrams \citep{mun96,yan96,mar07}.

We present, in Appendix~\ref{regss}, a schematic illustration of
these diagrams, and a brief description of how they were
previously interpreted by \cite{mun96}. We also extend the
analysis of these diagrams, including $V$-$\sigma$, to bring
additional insight to the observed bulk motions.

In Figure~\ref{int_sig} (left panel) we show the $I$-$\sigma$ plot
for the whole observed field. The plot for II Zw 40 shows exactly
the same basic features found in previous work on GH{\sc ii}Rs
\citep{mun96,yan96}. In general low intensity regions have the
highest $\sigma$ values while the range of $\sigma$ values
decreases for high intensity regions. In the case of II Zw 40,
there is a horizontal band defining a lower limit of $\sigma_{0}$
$\simeq$ 25 km s$^{-1}$ for regions brighter than 4000 counts (log F = 3.6).
 The horizontal line in the $I$-$\sigma$ plot shows the velocity dispersion
of 33.5 km s$^{-1}$ derived from the integrated line profile with
the 2.7$\arcsec$ aperture (Figure~\ref{feros} left panel). It is a
characteristic value for II Zw 40 found in all intensity regions.
The inclined bands identified by \cite{mun96} are also identified in
the plot. High velocity dispersions reach peaks of 65 km s$^{-1}$
in low intensity regions. As we will show in the next section, two
inclined bands in this plot are superposed in the range 2.6 $<$
log F $<$ 3.1.

Some characteristic line profiles are also shown  in
Figure~\ref{int_sig} (left panel). These are individual pixel
profiles exemplifying the variety of shapes observed 
in different regions. Profiles (a) and (b) are the most
asymmetric profiles observed. They originate close to the center
of the two regions of high dispersion at SE and at SW of the
brightest knot, respectively (see Figure~\ref{disp}). They are
clearly double-peak profiles, consistent with expanding motions.
Profile (c) originates in the second brightest knot and shows a
prominent red wing. Profile (d)  originates in the brightest
knot and profile (e) comes from the lowest dispersion region.


Figure~\ref{int_sig} (center panel) presents the $I$-$V$ plot 
which shows the relative radial velocity
variations. These motions are usually associated with the line broadening
component named $\sigma_{rad}$. The physical mechanisms at play 
which produce these line centroid variations
should be several, such as turbulence, winds, champagne flows,
expanding bubbles or rotation, acting on stellar radius scales
to hundreds of parsecs. However, the estimate of the contribution
 of these
subcomponents is limited to the spatial and espectral resolution
of the observations. For instance, in this work we chose to associate
the action of the spatially resolved bubbles, producing the
highest radial velocity variations, to the component
$\sigma_{rad}$. The other mentioned mechanisms, however, should
also contribute  to produce the overall pattern in the $I$-$V$
plot. Another line broadening component named
$\sigma_{shell}$ will be here associated with the unresolved
expanding shells identified as  responsible for
an underlying broad component found in the most inner region and
discussed in section~\ref{intprof}.
Therefore, the position-to-position motions found in
Figure~\ref{int_sig} (center panel) contribute to the total
broadening of an integrated line profile of the whole region. 
We do not intend to quantify precisely the contributions of
the broadening components from the diagnostic diagrams. In the
case of II Zw 40, the radial velocity differences are not higher
than 15 km s$^{-1}$ in the brightest central region, and the
highest differences are concentrated in low H$\alpha$ intensity
regions.
We found that the discrete tails toward high intensity values
shown in Figure~\ref{int_sig} (center panel) are associated with
the presence of knot A (main tail) and B with a relative radial
velocity of $\sim$ -13 km s$^{-1}$.

Figure~\ref{int_sig} (right panel) shows the
$V$-$\sigma$ plot. In the case of II Zw 40, it suggests that the
overall motions in the star forming region are primarily random,
instead of dominated by systematic line of sight motions.
For instance, if one was observing a highly collimated
outflow toward the line of sight, as a champagne flow, one should
expect to see a trend in this diagram. Broad profiles from the low
density regions should be found blueshifted, while narrow
profiles from high density regions should be found redshifted,
 producing a pattern such as a narrow
band with negative slope,  not found for the whole region
in Figure~\ref{int_sig} (right panel). On other hand,
the most important result shown by this diagram is that low
$\sigma$ (15-25 \kmsec) regions have small radial velocity
variations, suggesting that they are close to the rest frame of
the galaxy. This is the first indication that these regions may be
unperturbed by massive stellar evolution.




We will try to associate the morphological features observed in
the H$\alpha$ flux , velocity and dispersion maps with the features
from these diagnostic diagrams by choosing individual regions for
a more detailed analysis below.

\subsection{The Kinematics of Individual Regions Revealed by Diagnostic
Diagrams}\label{regs}

The patterns shown in $I$-$\sigma$, $I$-$V$ and $V$-$\sigma$ plots
(Figure~\ref{int_sig}) re-display many of the
features that can be observed in the  maps
(Figure~\ref{disp} and Figure~\ref{vel}). We wish to evaluate the
contribution of individual features to the integrated H$\alpha$
line profile of II Zw 40 using the kinematic diagnostic diagrams.
For this purpose, we defined eight regions with 
reliably measured $\sigma$ and radial
velocity.
The choice of regions was arbitrary, but motivated by different
peculiarities noted in the H$\alpha$ flux, radial
velocity and velocity dispersion maps. The chosen regions are
superposed on these maps in Figure~\ref{regions_map}.

\begin{figure*}

\epsscale{0.49} \plotone{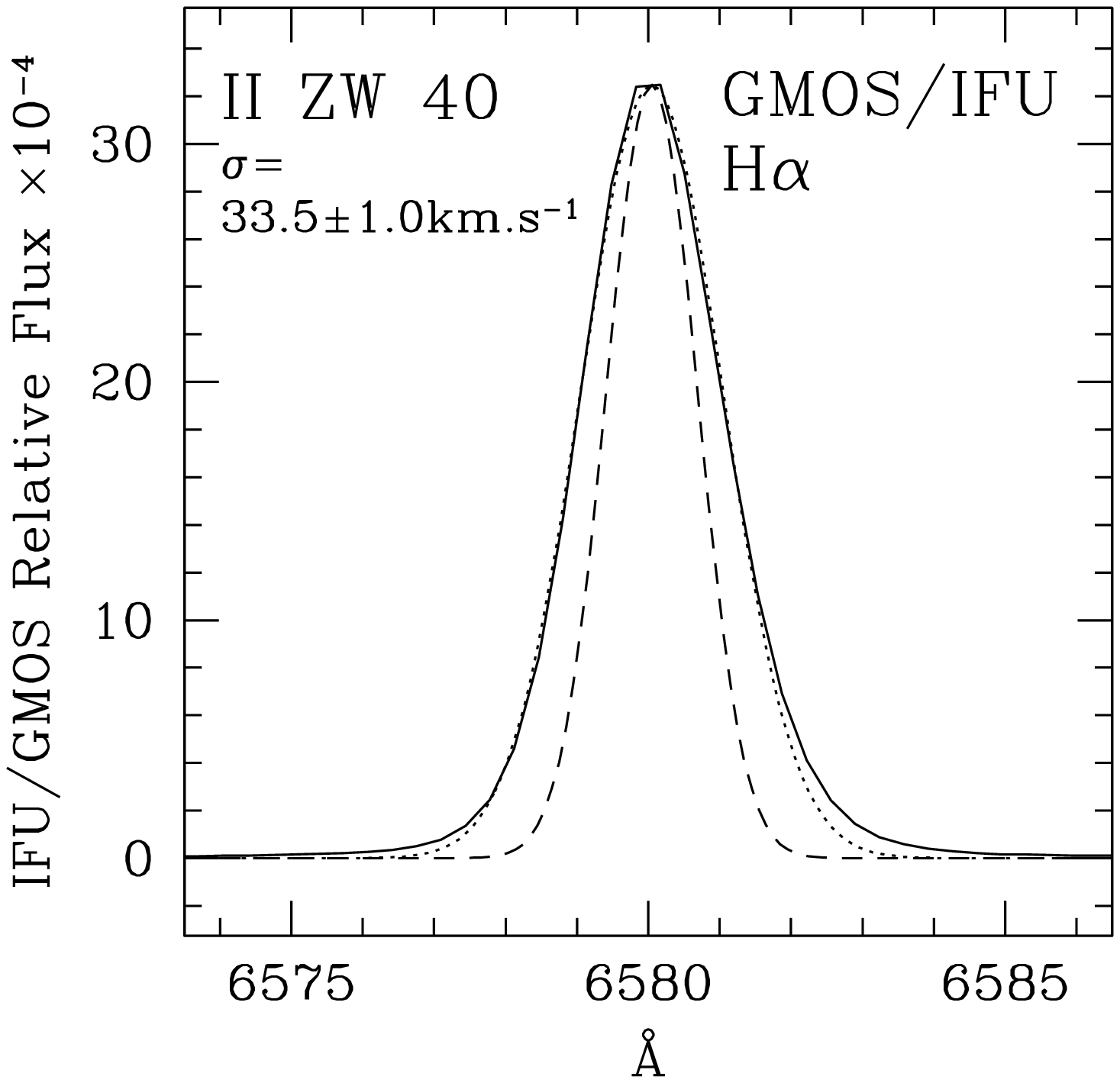}\plotone{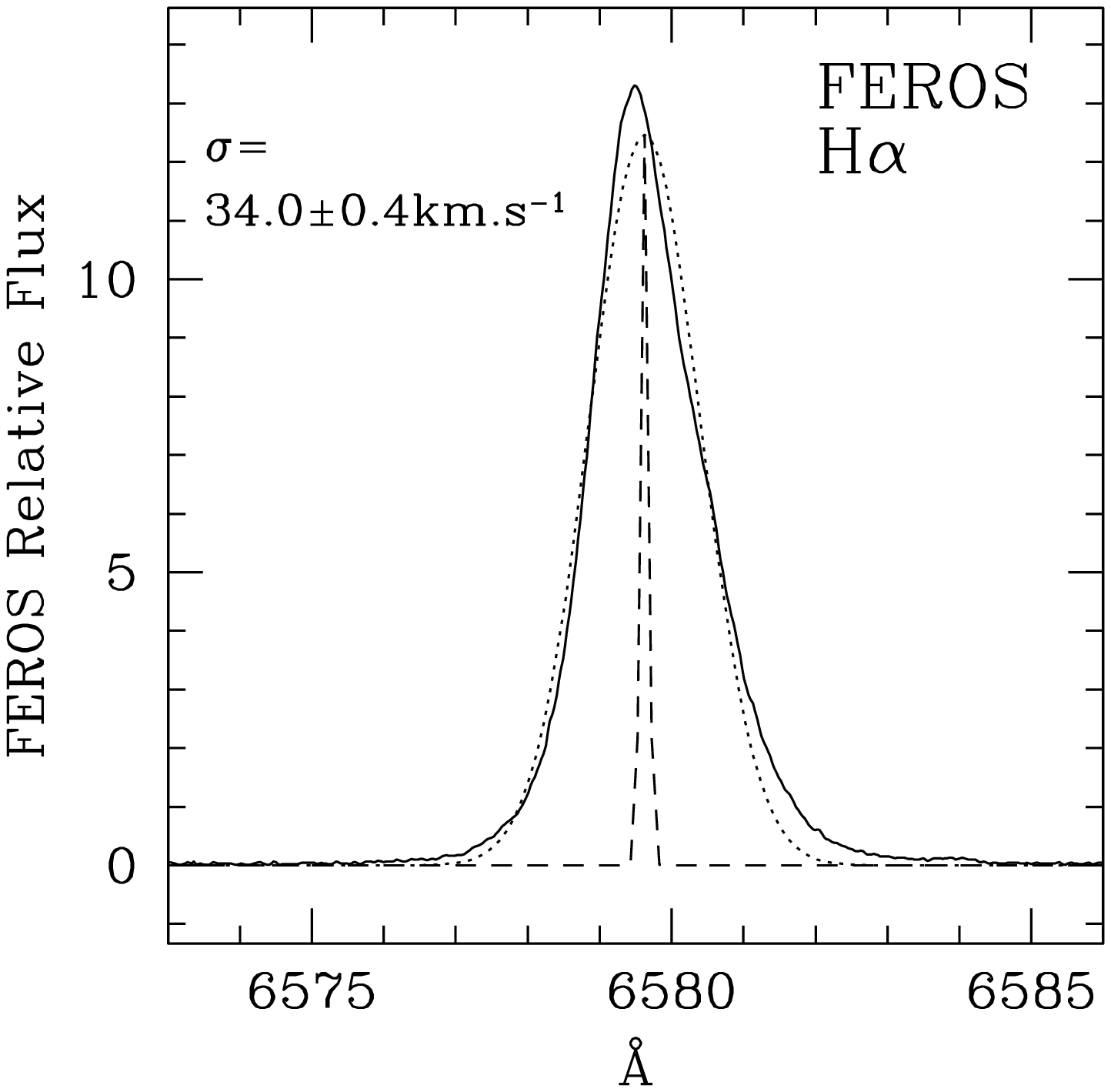} \caption{Line
profiles derived with GMOS/IFU (left) and FEROS (right) data. The
dashed lines represent the respective instrumental profiles (FWHM
= 1.5{\AA} for the GMOS and FWHM = 0.13{\AA} for the FEROS). The
dotted lines represent the single Gaussian fits. The velocity
dispersions values derived are shown in the boxes.\label{feros}}
\end{figure*}

\begin{figure*}
\epsscale{.90}
\plotone{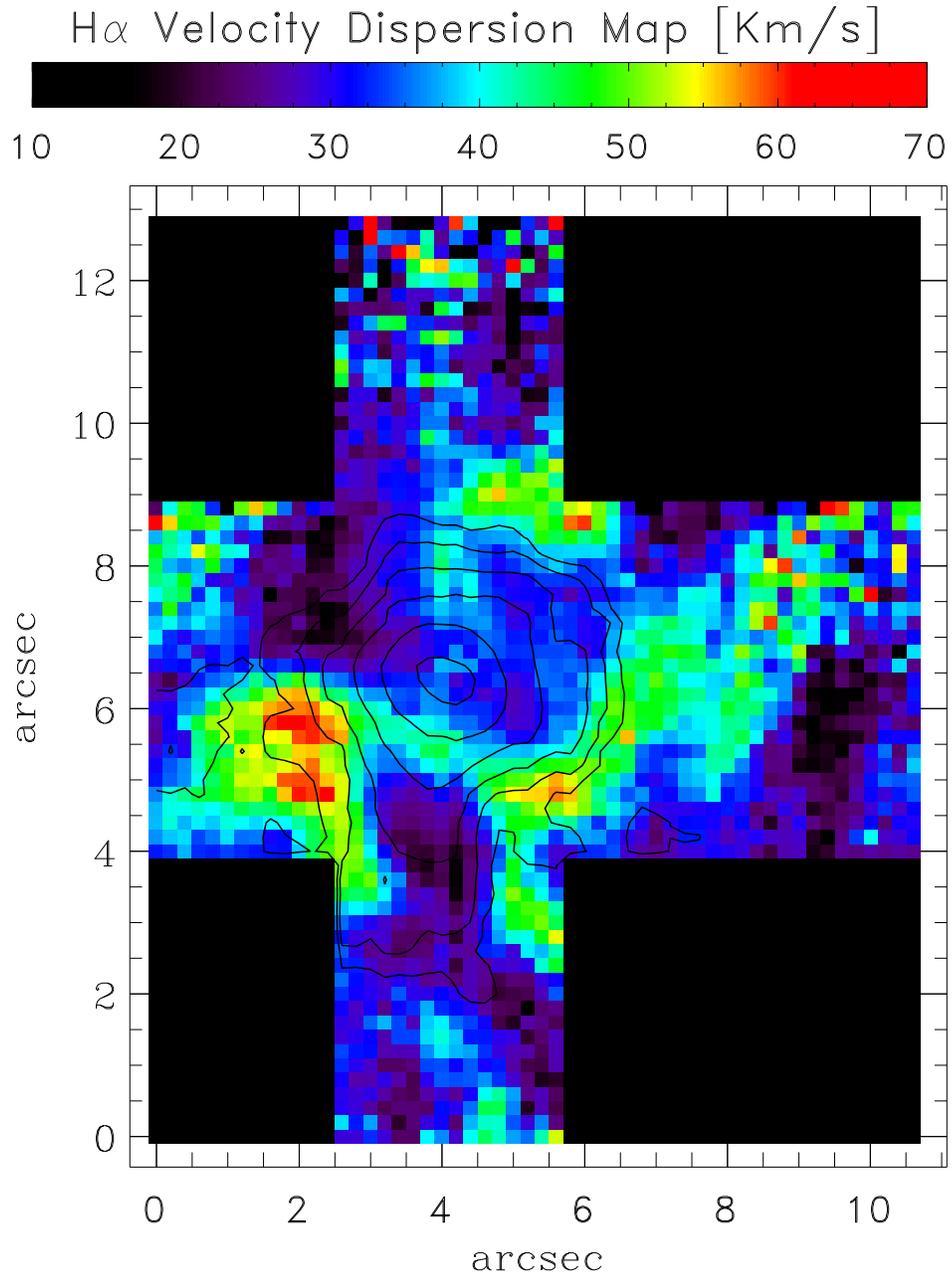} \caption{H$\alpha$ velocity
dispersion map, corrected for instrumental and thermal broadening,
 of the inner region of II Zw 40. The contours
represent 68, 20, 9, 5, 3.4 and 2.6\% of the H$\alpha$ peak
intensity.
North is up and east is left.\label{disp}}
\end{figure*}

\begin{figure*}
\epsscale{.90}
\plotone{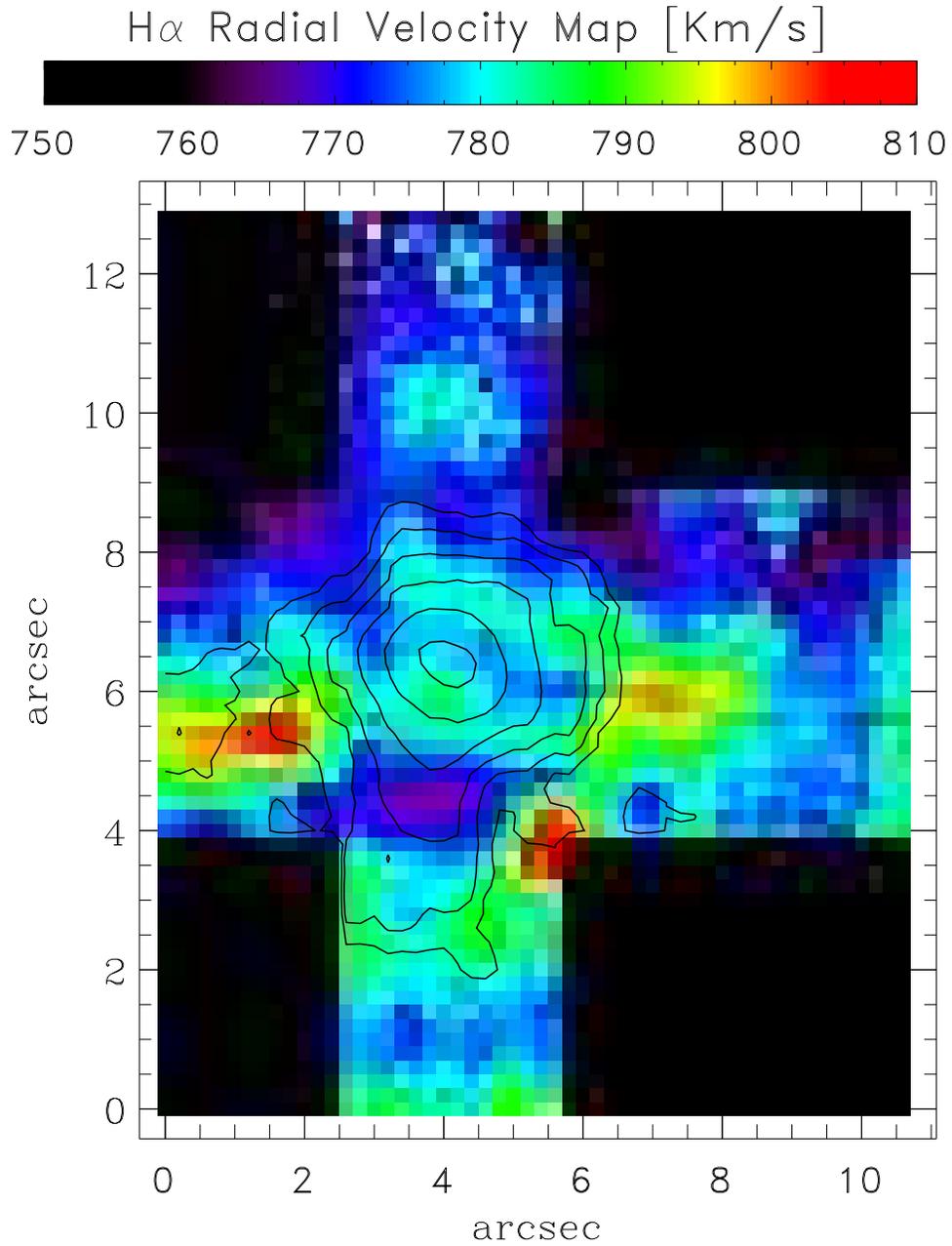} \caption{H$\alpha$ radial velocity
map. The contours represent 68, 20, 9, 5, 3.4 and 2.6\% of the
H$\alpha$ peak intensity. The recession radial velocity is not
corrected by the heliocentric motion. North is up and east is
left.\label{vel}}
\end{figure*}

For each chosen region, we have plotted a set of four
 graphs. These graphs are presented in
Figure~\ref{regions} in blocks of four boxes: (a) an integrated
profile over the region in linear-logarithmic axes, where we also
show the region number, the derived $\sigma$ value, and the
estimated skewness $\xi$ and kurtosis $\kappa$ of each integrated
profile, (b) $I$-$\sigma$, (c) $I$-$V$, and (d) the $V$-$\sigma$
plot. The vertical solid line in plots (b) and (c) shows our
confidence limit in flux as described in
Section~\ref{observations}. All references to the boxes in this
section refer to Figure~\ref{regions}, where we quote the region
number and the corresponding box letter, i.e. box (1a) means plot
(a) of Region 1 in Figure~\ref{regions}. The results of the
analysis of the emission line shapes through $\xi$ and $\kappa$
estimates for each region will be presented in
Section~\ref{intprof}. A brief description of each region follows:

{\bf Region 1} was defined to cover the main and brightest
H$\alpha$ knot (knot A) as shown by the H$\alpha$ map
(Figure~\ref{regions_map} left). It covers an area of 1.8$\arcsec$
$\times$ 1.6$\arcsec$ (equivalent to 100 pc $\times$ 90 pc). The
resultant integrated profile of Region 1 is shown in box (1a),
where the dashed and dotted lines represent the instrumental
profile and the Gaussian fit, respectively. The $\sigma$ derived
for Region 1 was 33.3 km s$^{-1}$ which agrees with the value
derived from our IFU simulated and FEROS spectra for a larger
aperture (see Figure~\ref{feros}). Box (1b) shows that there is a
well defined lower $\sigma_{0}$ value slightly above 25 km
s$^{-1}$ and no high $\sigma$ variations. Radial velocity
variations found in box (1c) indicates that the gas in Region 1
expands at $<$ 15 km s$^{-1}$. Since the points in box (1c) are
within $\sim$ 7.5 km s$^{-1}$ from the average recession velocity,
it is likely that the rms radial velocity variation contributes to
an equivalent FWHM of $\sim$ 15 km s$^{-1}$ or $\sigma$ $\sim$ 6.5
km s$^{-1}$ (see Appendix~\ref{regss}). Radial velocity variations
between individual pixels
in Region 1, therefore, do not contribute significantly to explain
the total 33 km s$^{-1}$ velocity dispersion. At this spatial
resolution only large scale radial velocity variations can be
identified, and only these will be analyzed here ($\sim$ 50 pc).
Box (1d) shows that most points are clustered with no apparent
correlation between V and $\sigma$. We found that random motions
are dominant in Region 1, instead of line of sight motions.

{\bf Region 2} was defined to cover the second brightest knot
(knot B), south of Region 1. Region 2 covers an area equivalent to
90 pc $\times$ 90 pc. Box (2a) shows that, the integrated profile
is the most asymmetric of all regions, showing a prominent red
wing. This region is responsible for the red wing in the
integrated profile obtained for II Zw 40 with FEROS, and also with
our GMOS/IFU data (Figure~\ref{feros}). The reason for this
asymmetry is that $\sigma$ reaches a local maximum in a few
relative high intensity points (North of knot B) as shown by box
(2b). Moreover, the same high intensity points, seen in box (2c),
have the largest recession velocities, resulting in this observed
red wing. As a consequence, box (2d) shows a strong correlation
pattern (see Figure~\ref{schem}C). The radial velocity motions in
Region 2 contribute twice as much to the integrate line width of
the region as compared with Region 1 ($\sigma$ $\sim$
13 \kmsec). The gas in  Region 2 has a
relative line of sight motion with respect to the gas in Region 1 of roughly 
-13 \kmsec. We found, however, two systematic components, one due to the
relative radial motion between the two knots, and the other due to
the expanding motion, as shown by the vertical band at log F
$\sim$ 3.2 in box (2c) presenting an intermediate radial velocity
variation.

{\bf Region 3} covers the SW high dispersion region. It covers an
area of 1.8$\arcsec$ $\times$ 1.6$\arcsec$ (equivalent to 100 pc
$\times$ 90 pc). The expected kinematic features due to the
presence of an expanding shell are seen in all boxes for this region. Box
(3a) shows a symmetric broad profile with $\sigma$ =
45.4 \kmsec. This is actually the resulting smoothed profile
obtained due to the contribution of unresolved asymmetries and
multiple components as shown in one pixel profiles (a) and (b) of
Figure~\ref{int_sig} (left panel). Box (3b) clearly shows the
inclined band indicating the presence of an expanding shell,
possibly produced by an
expanding bubble (see Figure~\ref{schem}A). Box (3c) confirms the
expanding motion through the very high spread in radial velocity
(see Figure~\ref{schem}B). The observed range of radial velocity,
of about 20 \kmsec~ above and below the average recession velocity should
underestimate the real radial component. This is in fact a
limitation of fitting single Gaussians to the integrated
profiles. From the line splitting in individual pixels
 in the center of the bubble we
could measure a maximum $\Delta V$ $\sim$ 90 km s$^{-1}$ or a
$V_{exp}$ $\sim$ 45 km s$^{-1}$. We
estimated the radial velocity flow component to the integrated profile shown in
box (3a) as FWHM $\sim$ 90 km s$^{-1}$ or $\sigma$ $\sim$ 38 km
s$^{-1}$ (see Appendix~\ref{regss}). Therefore, the expanding
motion  dominates the total broadening seen in the
integrated profile of Region 3. The data points in Box (3d) show
large scatter and there is not a significant trend in this plot,
as seen in Region 2.

{\bf Region 4} covers an area of 1.8$\arcsec$ $\times$
1.8$\arcsec$ (equivalent to 100 pc $\times$ 100 pc) and presents a
field of intermediate $\sigma$ and H$\alpha$ intensity values,
north of knot A. Region 4 presents many similarities with Region
1: similar $\sigma$ derived as shown in box (4a) and similar
pattern in box (4b) showing low $\sigma$ variations and a lower
limit $\sim$ 25 \kmsec. The small radial velocity variations found
in box (4c) and a concentrated velocity dispersion distribution in
(4d) indicate that there is no significant systematic motions
contributing to the total line broadening, which is dominated by
random motions on scales smaller than the seeing of $\sim$20 pc.

{\bf Region 5} is one of the most interesting regions, and 
covers a large area of 2.6$\arcsec$ $\times$ 2.6$\arcsec$
(equivalent to 150 pc $\times$ 150 pc)  SE of knot A. It
presents some of the broadest profiles, and it is probably
associated with an expanding shell (boxes 5a and 5b). Region 5
also presents high radial velocity spread (box 5c) of over
 40 \kmsec, although this may underestimate the real radial velocity 
component
as argued above in the analysis of Region 3. Regions 3 and 5 show
very similar patterns in all diagrams, suggeting that these
patterns can be easily used as expanding shell diagnostics in
GH{\sc ii}Rs. Despite their different sizes and morphologies the
two shell regions have similar kinematic properties. Some
double peak profiles seen in Region 5 are also consistent with a
bubble expanding with a velocity of 45 \kmsec, likely to be
powered by supernovae and stellar driven winds. There is not a
significant trend found in box (5d).

{\bf Region 6} was defined to cover a
field containing the lowest velocity dispersion values. It covers
an area of 2.0$\arcsec$ $\times$ 2.0$\arcsec$ (equivalent to 110
pc $\times$ 110 pc) North-East of Region 1. The observed $\sigma$
= 23.5 \kmsec, as shown in box (6a), is the lowest value verified
of all regions. Box (6b) also shows that a significant area
presents $\sigma$ values lower than  knot A, though they are still
supersonic. Velocity variations are also very small as shown by
box (6c). Box (6d) indicate primarily random motions. All plots
suggest, therefore, that Region 6 represents a region where the
mechanical energy input resulting from massive star evolution does
not play a significant role to stir the gas up.

{\bf Region 7} presents a relatively low velocity dispersion field
(box 7a). It covers an area of 1.6$\arcsec$ $\times$ 2.0$\arcsec$
(equivalent to 90 pc $\times$ 110 pc), South of knot B. It
completes the sampling of a larger low-to-intermediate dispersion
field covered partially by Region 2 (boxes 2b and 7b). Both
regions are confined between the two high dispersion bubbles,
Regions 3 and 5. The H$\alpha$ structure in Region 7 is associated
with the extended emission to the South of both knots (see
Figure~\ref{regions_map} left panel). Box (7c) shows a weak
correlation of $I$ and $V$ with fainter emission points moving
away. However, no clear correlation between radial velocity and
$\sigma$ is seen in box (7d).

{\bf Region 8} covers an area from W to NW of the knot A
(equivalent to 70 pc $\times$ 160 pc) with a $\sigma$ = 32
km s$^{-1}$ (box 8a), similar to
Region 1. It was defined to contain a field with almost constant
velocity dispersion through a large range of intensities, which is
shown by box (8b). The pattern formed in box (8b) is not
an inclined band. Region 8 does not present high
velocity variations, as shown by box (8c) and it is likely
dominated by random motion inferred by the points concentrated in
box (8d). The few higher dispersion points at low intensities seen
in box (8b) and (8d) may be picking up a possible additional
region of interest (i.e. shell) North of Region 8 (see
Figure~\ref{regions_map} right panel) which is not well mapped in
our observations and is not included in the present analysis.

Our definitions of the regions favor the interpretation of the
different patterns in $I$-$\sigma$ diagrams where kinematic
features are easily identified. Thus, the patterns may be the
result of different physical mechanisms. Something to have in mind
here, is our spatial resolution which only permits us to identify
kinematic features in scales of a tens of parsec and limited by
our field size. The overall picture is that local systematic motions
represent a dominant component to the observed line widths in
regions 3 and 5 which are likely associated with resolved
expanding shells due to the fact that we can observe line
splitting directly. Although, masked by the use of single
Gaussians, they leave a signature in the diagnostic diagrams.
Region 1 presents a small range of
intermediate $\sigma$ values which are also found in all the
intensities of nebular emission across the whole extent of the
star forming region. Region 2 is turbulent with some
characteristics of flow motions. Regions 4, 7 and 8 show smooth
turbulent fields. Wherever we cannot observe double peaked
profiles in these regions, we cannot completely rule out a radial
systematic component to the line width, possibly associated with
unresolved shells, however random motions must dominate over
expansion in these cases. Finally, relatively to the other
regions, Region 6 seems to be slightly affected or even
unperturbed by mechanical energy input provided by massive star
evolution, presenting the lowest $\sigma$ values.

\subsection{Making up the Integrated Velocity Profile from Individual Regions}\label{intprof}

Figure~\ref{diag_int} shows the fully integrated line profile
produced by combining all eight spectra for the regions presented
in \S~\ref{regs}, with a resulting $\sigma$ = 34.1 \kmsec. Despite
the complex medium-scale structures (50-100 pc) seen in the
synthetic maps (Figures~\ref{disp} and~\ref{vel}), mainly due to
the presence of asymmetries and multiple components (e.g.,
profiles (a) and (b) of Figure~\ref{int_sig} left panel), the
fully integrated profile is quite smooth and symmetric.

An important result is that the velocity dispersion derived from a
single Gaussian fit to the profile of Region 1
(Figure~\ref{regions} box 1a) is virtually the same as those
derived from the original FEROS spectrum with an aperture of
2.7\arcsec, and the simulated spectrum with the FEROS aperture on
our GMOS/IFU data (Figure~\ref{feros}), and the fully integrated
line profile (Figure~\ref{diag_int}). Region 1 represents 10\% of
the total area analyzed but 49\% of the total H$\alpha$ flux in
all eight regions. In contrast, Region 3 and 5, showing the
broadest profiles, cover 33\% of the total area analyzed but only
contribute 14\% to the total flux. Other regions, namely Region 6
and 7, covering 25\% of the total area and showing the lowest
$\sigma$ values only contributes 10\% to the total H$\alpha$ flux.
All the highest surface brightness regions, namely Region 1, 2, 4
and 8, show the common values $\sigma$ = 32-35 km s$^{-1}$
which characterize the global kinematics of the warm gas in II Zw
40's dominant starburst region.

The aperture effect to derive the characteristic velocity
dispersion of II Zw 40 is negligible if an integrated spectrum is
taken with an aperture covering the brightest knot. As an example,
\cite{mel88} derived for II Zw 40 a $\sigma$ = 35.2$\pm$0.5 km
s$^{-1}$ with a 6$\arcsec$ wide entrance slit aperture. From an
observational point of view, this result is very positive. If the
same is valid for most H{\sc ii}Gs and GH{\sc ii}Rs, as has
been confirmed, the measured supersonic $\sigma$, introduced in
the $L$-$\sigma$ relation, is little affected by the
effects of SNRs and wind driven bubbles where the warm gas
emission is faint, and where the integrated velocity dispersion is
much higher than in the central core.

We computed the skewness, $\xi$, and the kurtosis, or flatness
factor, $\kappa$, to quantify the shape of each profile and they
are included in all boxes (a) of Figure~\ref{regions}. We warn
that the quantitative results, especially for $\kappa$, are sensitive
to the window of summation and the height of continuum.
We defined a confident fixed window 6576-6584{\AA} to
compute the $\xi$ and $\kappa$ values, since broad wings are not
well mapped for the faintest profiles. Relative differences
between kurtosis values are more significant than their absolute
values, which should in principle be comparable with a Gaussian
profile ($\xi$=0, $\kappa$=3). We found that most profiles have
peaked shape $\kappa$ $>$ 3 (leptokurtic). Region 3 and 5 are
significantly more flattened than the others 
(platykurtic). The most asymmetric
profiles identified by the highest $\mid\xi\mid$ are those of
Region 2 and 6, which can be easily distinguished visually in
liner-logarithmic plots in boxes (2a) and (6a) of
Figure~\ref{regions}. We also estimated $\xi$ and $\kappa$ values
for the fully integrated profile shown in Figure~\ref{diag_int}.
We can see that not only the $\sigma$ but also $\xi$
and $\kappa$ values are virtually identical to those estimated
from profile of Region 1 (Figure~\ref{regions} box 1a).

Since the S/N in the wings of the profile from Region 1 is much higher
than the others, we were able to measure across in a larger
window (6574-6586{\AA}) and estimate another pair of $\xi$ and
$\kappa$ values for this profile, giving an idea of how sensitive
the values of the higher order moments are. The
estimated values derived in this spectral range are $\xi$=0.14 and
$\kappa$=4.19, showing that the integrated profile in
Region 1 is very peaked. This result indicates the presence of a
broad but weaker component in the core of II Zw 40. Multicomponent Gaussian
fits to the profile of Region 1 are shown in Figure~\ref{comp}. We
used the PAN routine (Peak ANalysis; \cite{dim05}) in IDL to fit two
components to H$\alpha$ profile. We found a narrow component
$\sigma_{n}$ = 29.4$\pm$0.6 km s$^{-1}$ which is not very
different from $\sigma$ derived from a single fit $\sigma$ =
33.3$\pm$0.9 km s$^{-1}$. The broad component was estimated to be
$\sigma_{b}$ = 73.2$\pm$1.0 km s$^{-1}$ and its flux corresponds
18\% to the flux of the narrow component. We may speculate that
the broad component
is associated with unresolved wind-driven shells at the
core of II Zw 40 where most of the young and massive stars
concentrate.

\section{Discussion}\label{discussion}

An important verification provided by the above results is that
the H$\alpha$ kinematic features already found in Local Group's
GH{\sc ii}Rs are represented in the main starburst region of II Zw
40: a prototypical H{\sc ii} galaxy at roughly 11.9 Mpc. In fact,
one way to investigate the starburst phenomena on these galactic
scales is to compare the main starburst region in II Zw 40 with
the local GH{\sc ii}Rs  studied by similar techniques.

\cite{yan96} found that virial motions and expanding shells
contribute roughly equally to the velocity width of the integrated
profile in NGC 604. Using the same methods, \cite{mun96}
proposed an evolutionary scenario when they compared their results
of NGC 604 with NGC 588. A simple model for an idealized
homogeneous shell evolving in a uniform density medium suggests
that younger and fast shells should have lower intensities and
 reach higher $\sigma$ values than older ones. Hence, the
authors argue that NGC 604 appears to be younger than NGC 588
based on the inclined bands associated with shells, and identified
in $I$-$\sigma$ plots. Furthermore, NGC 604 presents a well
defined lower limit ($\sigma_{0}$ $\simeq$ 17\kmsec) in supersonic
line width everywhere (horizontal band in $I$-$\sigma$ plot),
while NGC 588 does not. A substantial fraction of NGC 588 presents
line widths lower than the brightest region, including points
showing subsonic $\sigma$ values. \cite{mel99} suggest that it is
also the case for 30 Dor. In the Cometary Stirring Model (CSM)
framework, this difference can also be understood as an
evolutionary effect. A young region keeps most of the supersonic
motions communicated to the leftover gas via bow shocks by the
winds of low-mass stars during its early stage. As massive stars
evolve, ejecting mechanical energy into the ISM, through stellar
winds and supernova events, they produce shells that expand beyond
the core. Shells will be destroyed by random motions of
the background neighboring gas and slow down, resulting in  $\sigma$
even below the bright region \citep{ten93,mun96}. The main point
in this scenario is that the dominant line-broadening mechanism in
young GH{\sc ii}Rs is gravity, while tuburlence by wind-driven shells 
and supernovae dominate
in older ones.

More recently, \cite{mar07} have shown that the morphological
features observed in their analysis of three-dimensional
spectroscopic data of three blue compact galaxies (Mrk 324, Mrk
370, and III Zw 102), especially those in the $\sigma$ $versus$
$I_{peak}$ diagram, look similar to the ones found in GH{\sc
ii}Rs, with a supersonic horizontal band that extends over a large
range of intensities, and inclined bands that reach high $\sigma$
values at low intensity arising from large volumes surrounding the
central knots. In a similar study to our own, using similar
technique with Gemini GMOS/IFU, \cite{wes07a,wes07b} looked at the
properties of a young star cluster  and its environment in the
dwarf irregular starburst galaxy NGC 1569. Despite the fact that
they decompose the profiles rather than fit a single Gaussian, the
narrow components of their profiles show all properties found in
these other studies including ours, namely that, it is most likely
explained by a convolution of the stirring effects of massive star
evolution and gravitational  motions. In addition, they
found that the observed lower limit ($\sigma_{0}$ $\simeq$ 12 km
s$^{-1}$) of the narrow component is in agreement with
$\sigma_{virial} = \sqrt{GM/R}$.

Here, we wish to test the hypothesis that the dominant broadening
mechanism of the H$\alpha$ line in the core of II Zw 40 is
gravity.

As shown in Figure~\ref{int_sig} (left panel), II Zw 40 presents
the lowest $\sigma$ values ($\sigma$ $\sim$ 20 km s$^{-1}$) in
regions with faint H$\alpha$ emission. Here we identify Region 6
as the most significant contributor to this low $\sigma$ (see
Figure~\ref{regions} box 6b), which appears to present dominantly
random motions, through its diagnostic
diagrams. In addition, Region 6 does not seem to be affected by 
the neighboring superbubble of Region 5.
Therefore, this region may still retain the kinematic information
of the pre-existing underlying turbulent velocity field.
The velocity
dispersion values derived in Region 6 should not be associated
with the internal velocity dispersion of individual dense
molecular cloud cores or clumps (typically less than a few \kmsec),
where individual or groups of massive stars form, but
instead, they may be the result of the large scale turbulent
motions of the parent complex of diffuse molecular clouds, which
gave rise to the present episode of violent star formation. Some
of these clouds will form stars in short local dynamical times,
while others, though dense, may not form stars. We do not intend to
discuss any detailed model, but simply try to envisage a viable
scenario for our conjecture of an overall underlying velocity
component. Massive star clusters will form in much denser
fragments (clumps or cores) of the molecular clouds whose
properties are not assessed by our observations, and require
density tracers (e.g. HCN) with high spatial resolution radio
observations. Though, these clumps must also have relative velocities
stirring up the ISM under the influence of the common
gravitational potential due to the complex of stars, and gas, and
possibly dark matter. The $^{12}$CO (J= 1 $\rightarrow$ 0) traces
the diffuse molecular cloud on hundreds of parsec scales. The
measured CO line width of a region encompassing the entire ionized
region in II Zw 40 is $\Delta$V = 42 km s$^{-1}$ or $\sigma$
$\sim$ 18 km s$^{-1}$ \citep{tac87}, somewhat comparable to the
low $\sigma$ in the ionized gas. However, these motions seem to be
completely detached from the global HI velocity distribution
which extends up to 5 arcmin ($\sim$ 6 Kpc) in the direction of
the southern and northern tails \citep{van98, bri88}, where the
line width is W$_{50} \sim$ 120 km s$^{-1}$ ($\sigma \sim$ 50 km
s$^{-1}$) and has a strong contribution by the large scale
rotation pattern of the galaxy on these scales.

The large resolved supperbubble in Region 5 is very well delimited
in $\sigma$ and intensity maps and may not have slowed
down sufficiently to be disrupted. Moreover, Region 3 also
presents a well defined bubble in $\sigma$ and intensity maps,
which is seen as an inclined band in the diagnostic plot. Despite
being smaller than the bubble in Region 5, they both show
similar profile shapes, expansion velocities $V_{exp}$ $\sim$ 45
\kmsec, and integrated velocity dispersions $\sigma$ $\sim$ 45
\kmsec, possibly indicating similar evolutionary phase and
associated with the age of knot B.

Although II Zw 40 contrasts slightly with a simple scenario for a
GH{\sc ii}R, in the sense that it may suffer from the composite
effects resulting from the evolution of multiple stellar clusters,
we can say that in the CSM context, the GH{\sc ii}R in II Zw 40 is
kinematically young. Some superbubbles are evolving within the
core, though it does not present a constant
$\sigma$ or only ``well-behaved'' Gaussian emission lines
\citep{mun95,mun96}. However, the presence of surrounding regions
showing lower $\sigma$ than the core in II Zw 40 is a sign of
youth. There are still unperturbed regions preserving kinematic
information of the parent cloud under the dominant influence of the
galaxy's gravitational potential, ionized by diffuse UV radiation.
In consonance with our conclusions, many studies about this galaxy
indicate the presence of a very young starburst in its brightest
region, as mentioned in the introduction.

An important resolution bias must be of concern here in order to
test our hypothesis. \cite{mel99} show that the supersonic
integrated line profile in 30 Dor is actually due to a
superposition of discrete number of clouds, small shells and
filaments which are resolved on subparsec scales. In these scales,
strong winds of WR, O and B stars might be responsible for
stirring up the gas around the dominant young cluster, imposing a
turbulent motion higher than the one predicted only by the
gravitational potential. The presence of a significant number of
the very young stars in the core should be responsible partially
for the difference between the $\sigma$ found in Region 1 and that
one found in Region 6. In fact, we found evidence of a broad
component $\sigma_{b}$ $\sim$ 73 km s$^{-1}$ in Region 1,
responsible for the large wings in the integrated profile, which
may be associated with unresolved inner shells. However, its
contribution to the integrated profile is very small, since only
18\% of the total H$\alpha$ flux in Region 1 comes from this
component.

We can now sum up the main reasoning so far. A common framework
assumes that the observed velocity dispersion ($\sigma_{obs}$) is
mainly given by the contributions of gravity ($\sigma_{grav}$),
thermal broadening ($\sigma_{th}$), multiple unresolved expanding
shells ($\sigma_{shell}$) and systematic radial motions
($\sigma_{rad}$). Here we consider $\sigma_{rad}$ as being the sum
of several phenomena such as SN, expanding bubbles, champagne
flows, relative motions between discrete clusters providing large
scale shear, turbulence and rotation, which cause
position-to-position radial velocity variations and, in extreme
cases, line splitting. We detected and identified $\sigma_{shell}$
only in the core, i.e. Region 1, as the weak broad component, but
possibly present with different weights in the whole nebula.
Indeed, we found that most profiles have peaked shape indicating
the presence of faint wings or broad components, with the exception of
those found in regions of bubbles, i.e. Region 3 and 5. Once
corrected for the instrumental profile, the observed velocity
dispersion of a GH{\sc ii}R or discrete regions inside them is
therefore given by

\[\sigma_{obs} = \sqrt{\sigma_{grav}^{2} +
\sigma_{shell}^{2} + \sigma_{th}^2 + \sigma_{rad}^2}\]
\citep{mel99}.
We fixed the thermal broadening as $\sigma_{th}$ = 9.1 km s$^{-1}$
due to the gas electron temperature for hydrogen at 10$^{4}$K,
causing no significant systematic error in our results. The radial
velocity variations on scales larger than 50 pc can be identified
here. It varies from few km s$^{-1}$ in regions such Region 1,
$\sigma_{rad}$ $\sim$ 6.5 km s$^{-1}$, to $\sigma_{rad}$ $\sim$ 38
km s$^{-1}$ in Regions 3 and 5 as inferred from line splitting in
the center of bubbles. Region 6 seems to be
 ``free from shells'', and it can be a good
probe of $\sigma_{grav}$ $\sim$ 20-25 km s$^{-1}$, posing a lower
limit of velocity dispersion in II Zw 40. In fact, these values
are not very different from the narrow component found in Region
1, $\sigma_{n}$ $\sim$ 29 km s$^{-1}$, and the lower limit
$\sigma_{0}$ $\sim$ 25 km s$^{-1}$ in the $I-\sigma$ diagnostic diagram.
If indeed, the $\sigma$ measured in Region 6, is unaffected by
stellar evolution, and the narrow component of the Gaussian fit
to the profile of Region 1 gives the same low value of $\sigma$,
and additionally the low limit of overall measured $\sigma$ is probing the same
physical mechanism, we may
suggest a common dominant line broadening cause.
Gravity  provides the continuous underlying source of stirring the ISM 
over the whole star-forming region.
However, the characteristic $\sigma$ = 32-35 km s$^{-1}$ found over the whole
starburst region of II Zw 40 and derived from a single Gaussian fit should
be the result of combined sources, of the dominant action of gravity, unresolved multiple
expanding shells and systematic radial motions.

\section{Summary}\label{summary}


The three-dimensional spectroscopic analysis of II Zw 40, using
Gemini-North GMOS/IFU,
yield insights into several astrophysical questions about the
structure and evolution of starburst regions, specially those
associated with the birth of super star clusters and their impact
on the ISM. The kinematics of the warm gas inferred from line
profiles of emission lines, such as H$\alpha$, imposes constraints
on the theoretical models of star formation
and on feedback of stellar evolution on the ISM. We have
analyzed the kinematic properties of the ionized gas in II Zw 40
using diagnostic diagrams and individual region
description. A summary of the main conclusions follows:

\newcounter{bean1}

\begin{list}
{(\arabic{bean1})}{\usecounter{bean1}}

\item Electron densities in the inner starburst region range from
$5\times10^2$ to $1.4\times10^3$ cm$^{-3}$, somewhat higher than
the common low density regime for H{\sc ii} regions of 10$^{2}$
cm$^{-3}$.  Its spatial distribution is not constant and shows
a clear decline (compatible with an r$^{1/4}$ law) with a
peak coincident with the H$\alpha$ line emission peak.

\item Oxygen abundance as inferred by the N2 calibrator in the
inner 100 $\times$ 100 pc$^{2}$ ranges from 12+log(O/H) = 7.8 to
8.4, showing lower values 12+log(O/H) = 7.85 associated with
the  dominant site of star formation, namely knot A.

\item The aperture effect to infer the characteristic velocity
dispersion $\sigma$ = 34 km s$^{-1}$ of II Zw 40, derived from
a single Gaussian fit, is negligible. All shape properties
($\sigma$, $\xi$ and $\kappa$) of a fully integrated profile are
virtually the same as those found in the profile of the brightest
region.

\item The diagnostic diagrams of $I$ $versus$ $\sigma$, $I$
$versus$ $V$, and additionally $\sigma$ $versus$ $V$ are powerful
tools to identify the origin of internal motions in GH{\sc ii}Rs
and also in H{\sc ii}Gs.

\item We found regions where stellar evolution does not seem to
play a significant role  stirring the gas up, and should reveal
the kinematic signature of the proto-cloud that gave rise
to the present starburst. In the case of II Zw 40, Region 6
North-East of the main knot poses a lower limit of $\sim$ 23
\kmsec~ to the global velocity dispersion in the galaxy, which
should provide a good estimate of $\sigma_{grav}$.

\item Despite the different sizes and morphologies, the two large
shells found in Region 3 and 5 have the same kinematic
properties, with very similar patterns in all diagnostic
diagrams. Both regions have smoothed integrated line profiles
with $\sigma$ $\sim$ 45 \kmsec~ and they seem to expand at
$V_{exp}$ $\sim$ 45 \kmsec, contributing significantly with
$\sigma_{rad}$ $\sim$ 38 \kmsec~ to the total width of their
integrated profile.

\item The inner 50 pc and brightest region of II Zw 40 seems to be
kinematically very young, where gravity seems to play an important
role. Part of the difference found between $\sigma$ $\sim$ 33
km s$^{-1}$ in the core and 23 \kmsec~ found in Region 6 is
possibly associated with a weak broad component and a
small systematic radial velocity flow component $\sigma_{rad}$.
The broad component should
be associated with $\sigma_{shell}$ originated from multiple
unresolved shells provided by stellar winds of WR and OB stars
acting on few or subparsec scales.

\item The kinematic features in II Zw 40 are all remarkably
similar to the ones found in supersonic GH{\sc ii}Rs in irregular
and other star forming galaxies powered by massive star and stellar 
cluster formation and
evolution,  suggesting that these starbursts in HII galaxies
are  their scaled up versions on galactic scales. 

\end{list}

Finally, the core supersonic $\sigma$ is the one producing the
Luminosity-$\sigma$ relation observed in GH{\sc ii}Rs and H{\sc
ii}Gs. This
component can be derived by simply measuring the integrated line
profiles by fitting a single Gaussian on the brightest knot of the
starburst, regardless of aperture size. However, we must further
investigate how these derived motions precisely relate to the
underlying mass distribution before we can derive absolute total
galactic masses. We must also investigate, with a statistically
significant sample, the possible evolutionary effects of the
starburst on the observed relations, and how they can be
parametrized for the use as a powerful extragalactic distance
estimator applied to high redshifts.




\acknowledgments VB acknowledges the CAPES grant from the
Brazilian Federal Government. HP acknowledges CAPES financial 
support under process number BEX 3656/08-0.
ET acknowledges his US Gemini Fellowship by
AURA.  We would also like to thank Jorge
Melnick, Alberto Bolatto and Zhi-Yun Li for fruitful discussions.
We are specially  thankful to Mark Whittle for carefully
reading a revised version of the manuscript and
providing critical comments, as well as English proofreading.
Finally, we appreciate the comments, criticisms and suggestions 
by the referee which significantly improved the paper. 


\appendix

\section{Interpreting the Diagnostic Diagrams}\label{regss}

\begin{figure*}
\epsscale{0.32} \plotone{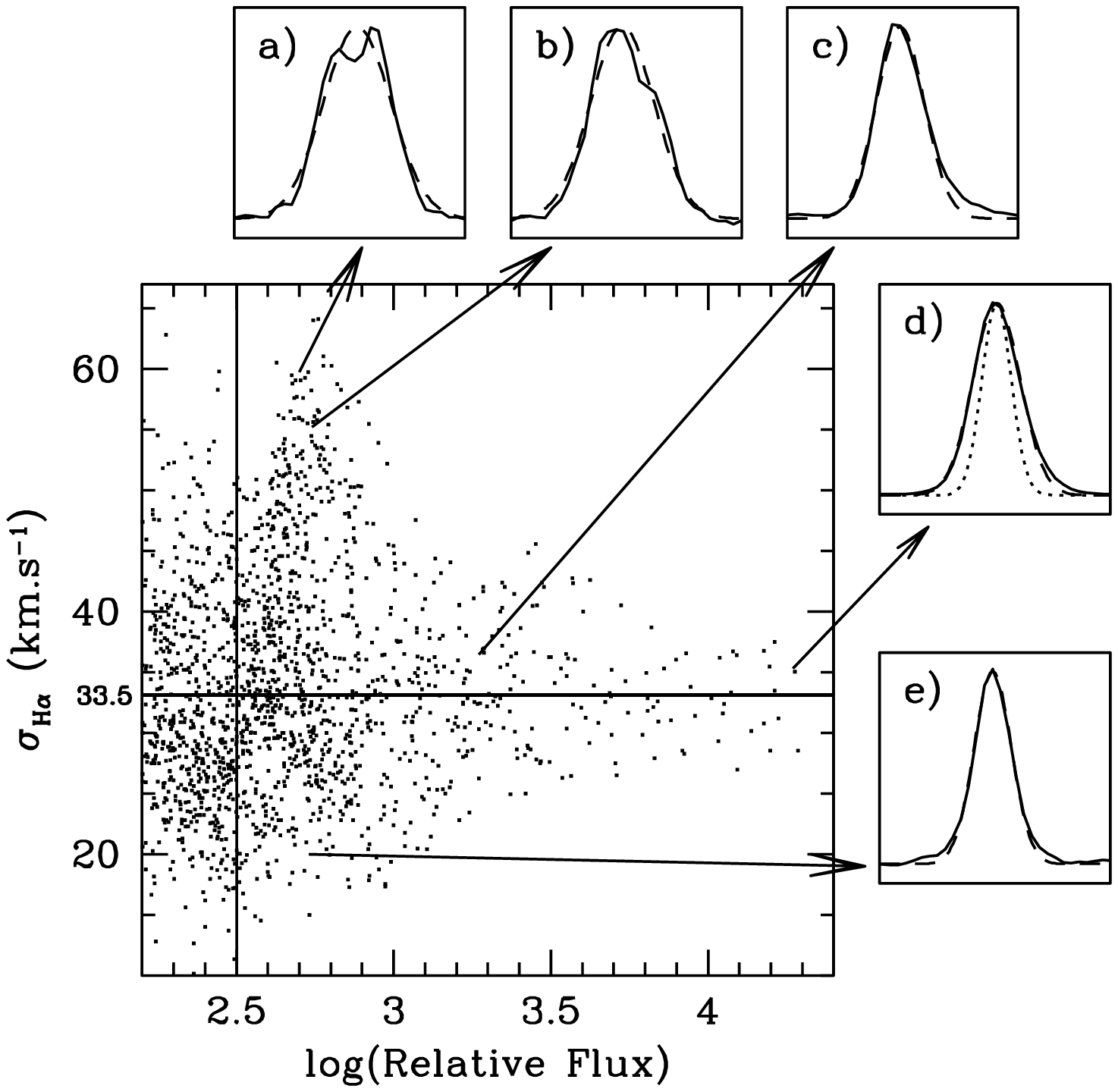}
\plotone{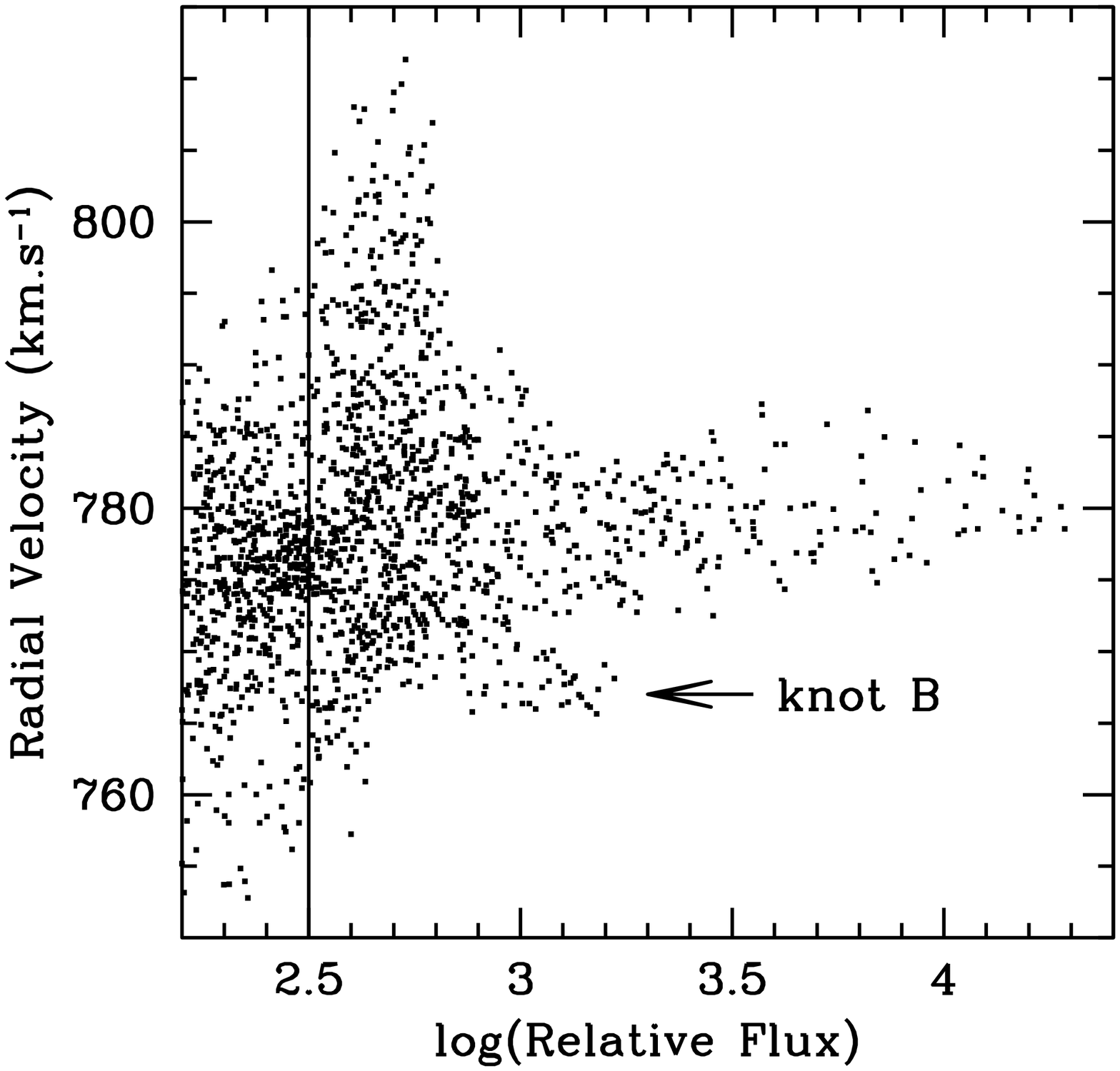}\plotone{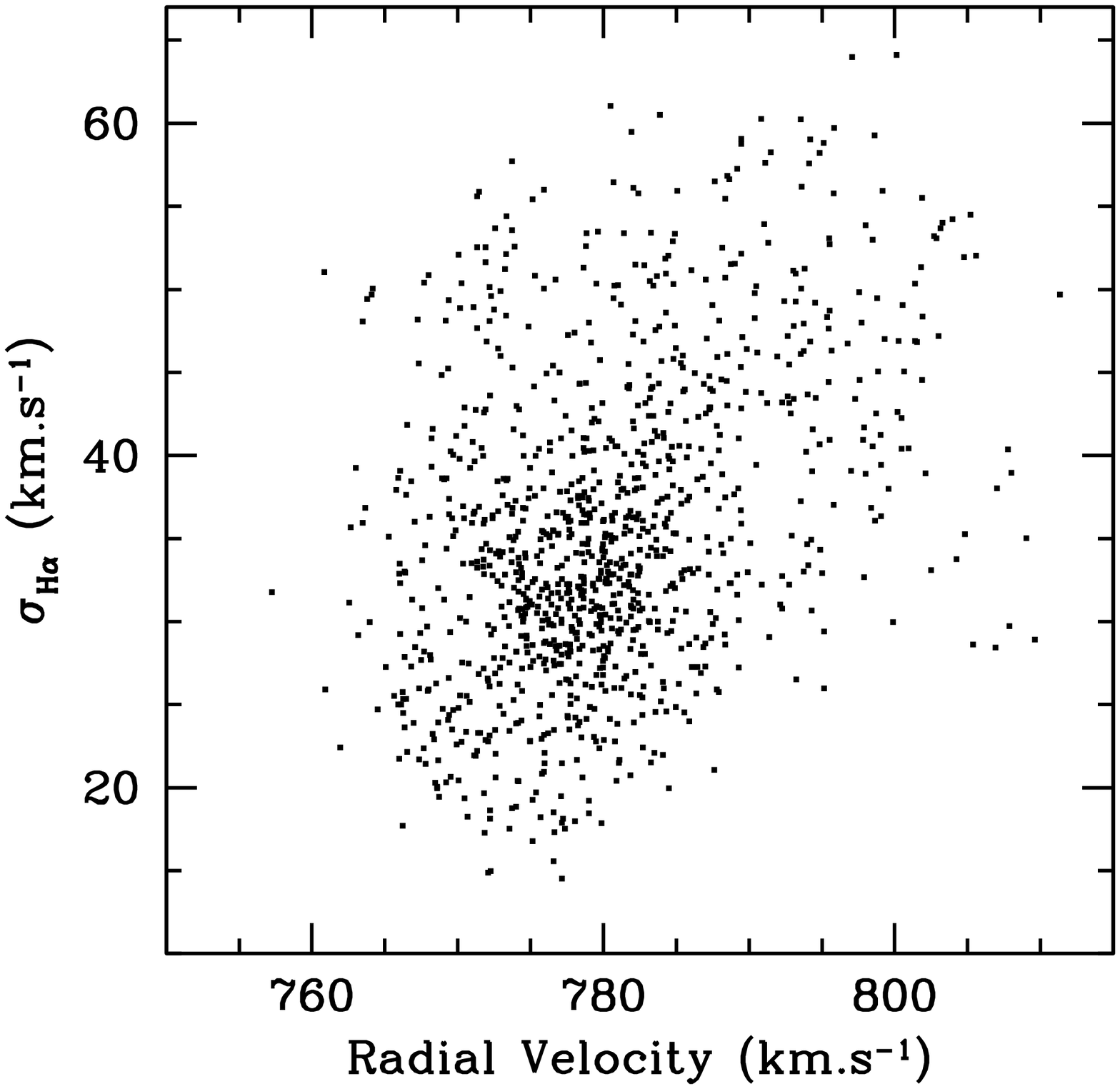}\caption{$I$-$\sigma$ (left),
$I$-$V$ (center) and $V$-$\sigma$ (right) plots for the whole
region observed in II Zw 40 as described in the text. The solid
horizontal line in $I$-$\sigma$ plot represents the $\sigma$ value
for the inner core (D = 2.7$\arcsec$). Characteristic individual
pixel line profiles are shown in the left panel from (a) to (e).
Dashed lines represent the single Gaussian fits and the dotted
line under profile (d) is the instrumental profile. The vertical
solid line at log(Relative Flux) = 2.5 represent the estimated
minimum line flux to well measure a single line profile. All
points with log(Relative Flux) $<$ 2.5 are not shown in
$V$-$\sigma$ plot.\label{int_sig}}
\end{figure*}

\begin{figure*}

\epsscale{0.32}\plotone{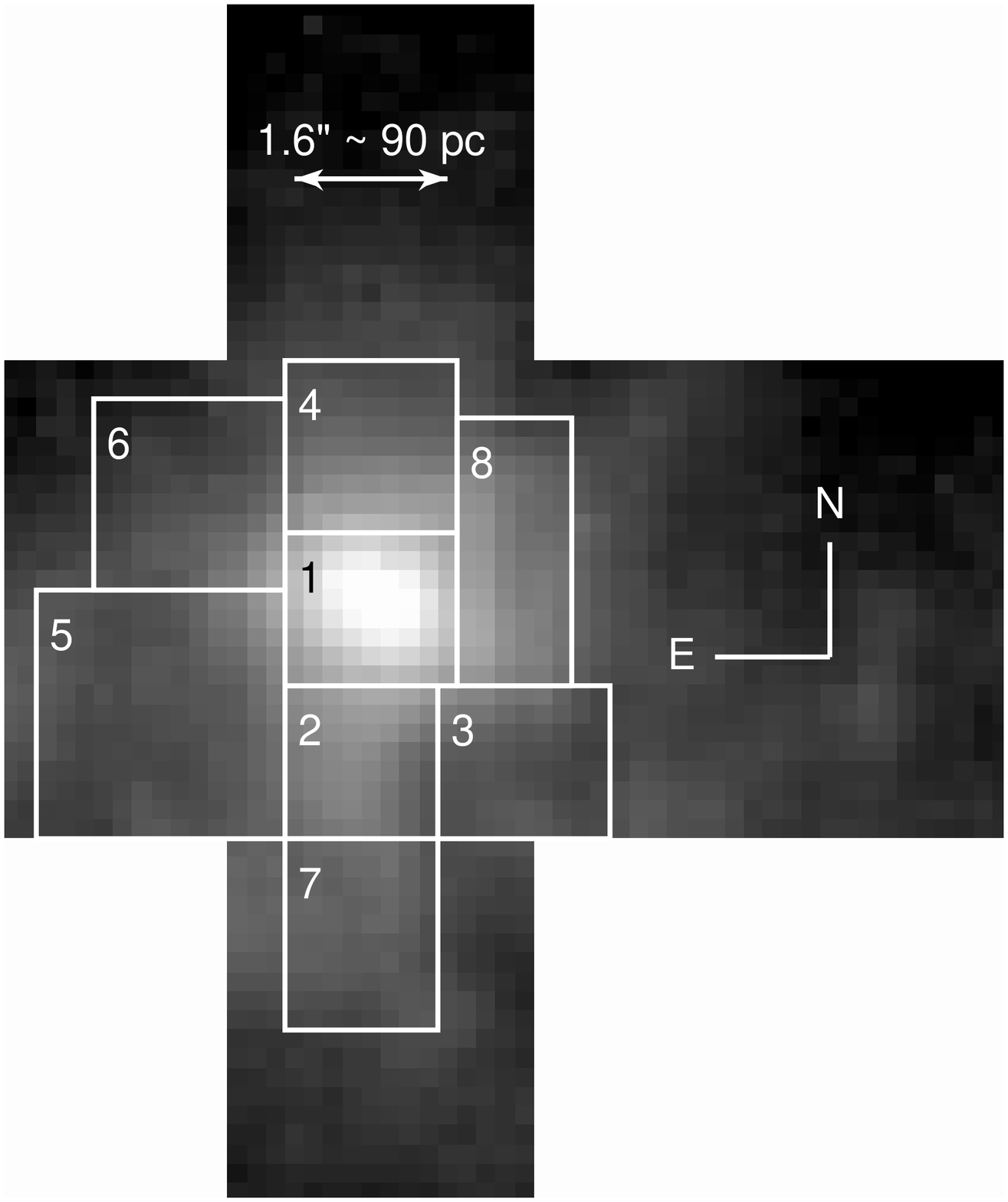} \plotone{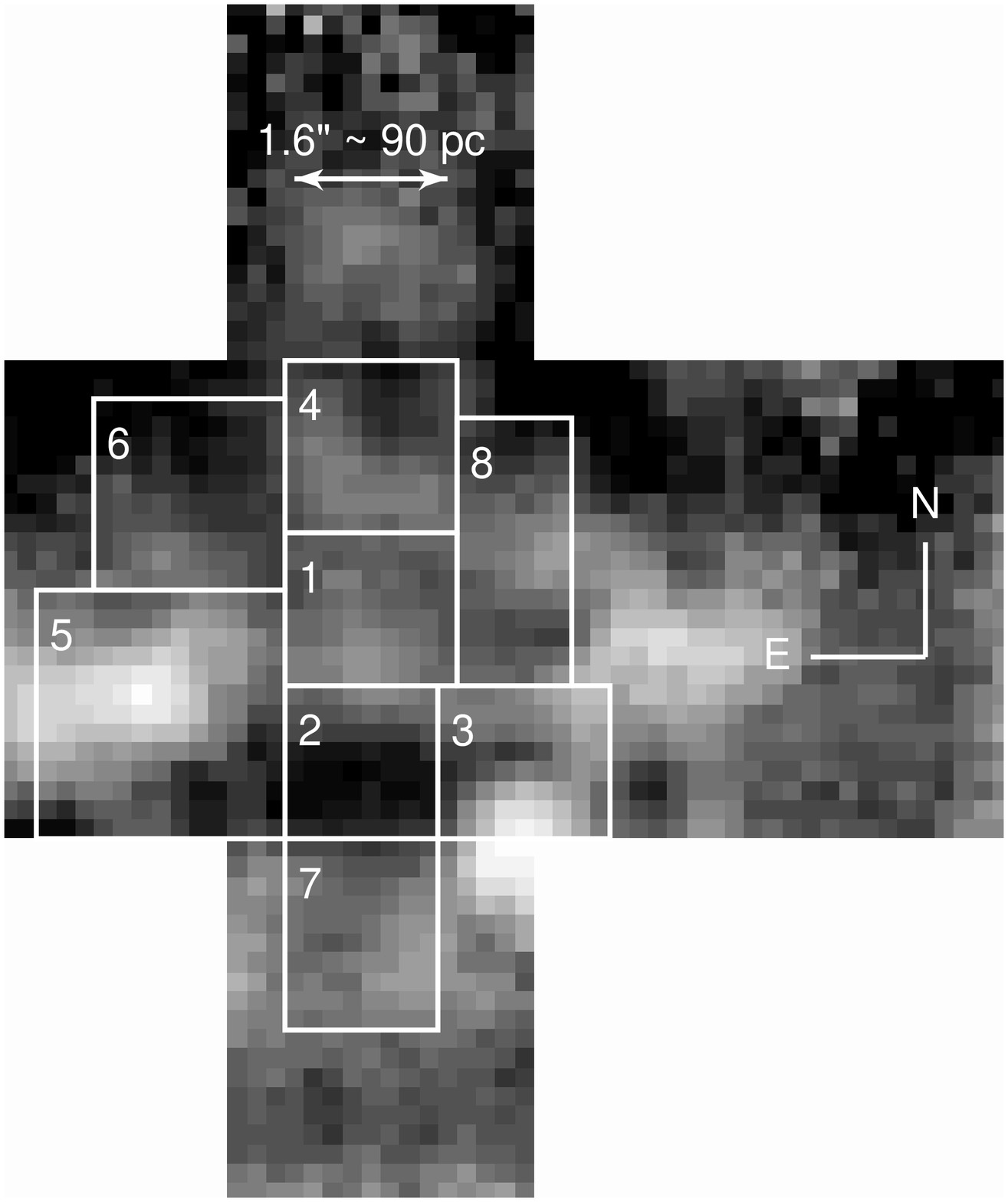}
\plotone{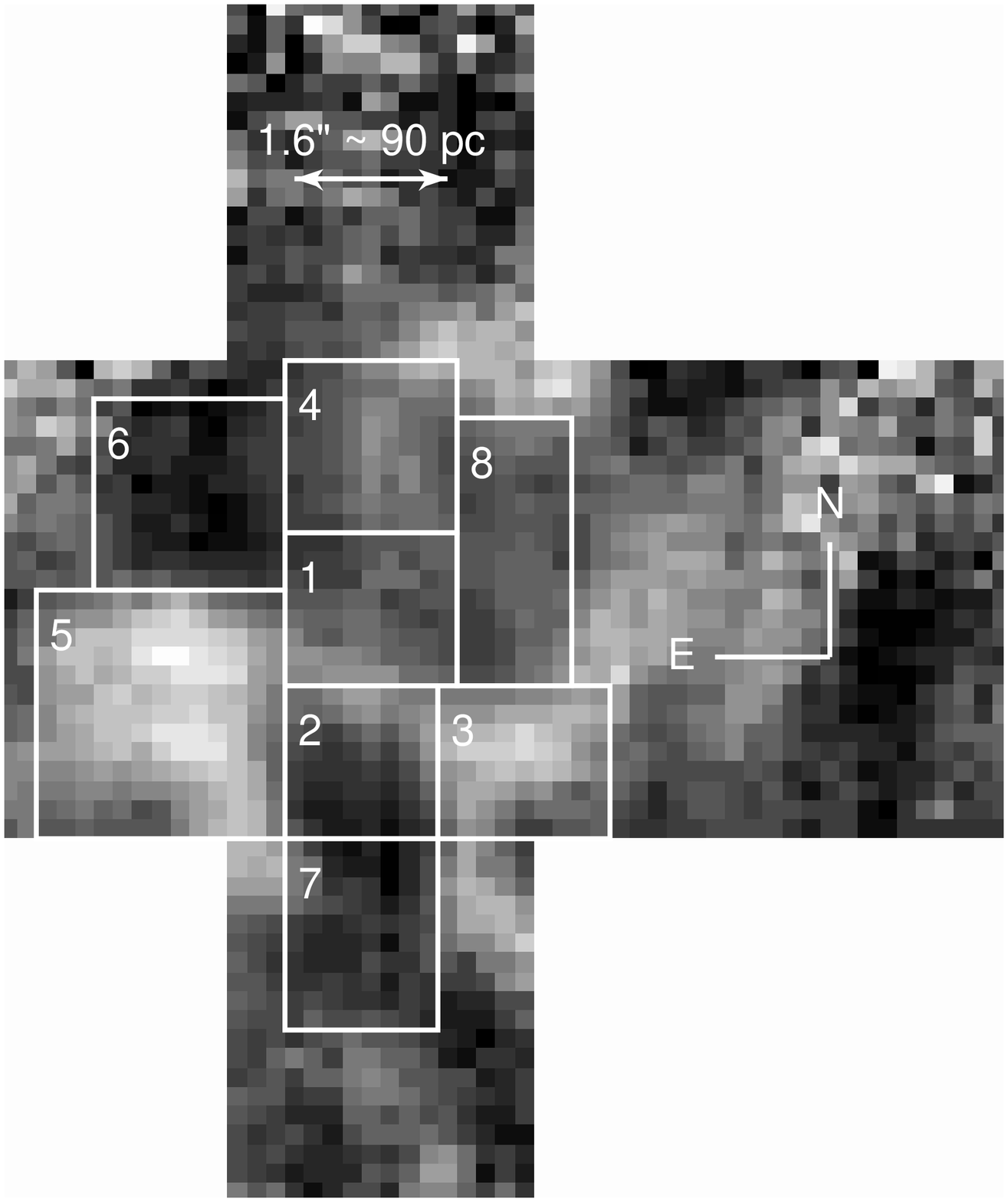}

\caption{H$\alpha$ emission map (left), radial velocity map
(center) and velocity dispersion map (right). The eight regions
defined are drawn on each map.\label{regions_map}}
\end{figure*}

It is convenient to emphasize the meaning of some patterns formed
by the point distributions in the different diagnostic diagrams of
Figure~\ref{int_sig}. These patterns are the graphical
representation of the kinematic signatures observed in the 2D
flux, velocity and dispersion maps. The following
description is valid for the methodology of fitting only single
Gaussians to the emission line profiles and its limitation is
emphasized when necessary. However, it is a very powerful tool for
kinematic analysis provided by data cubes. Three typical patterns
are of particular interest, and they are sketched in diagrams
(A), (B) and (C) of Figure~\ref{schem}.

Figure~\ref{schem}A shows the pattern known as the inclined band
in the $I$-$\sigma$ diagram, previously identified by
\cite{mun96}. The authors propose that this pattern should be the
result of an ideal shell evolving in the ISM. Their scenario
allows the interpretation of this diagram as a diagnosis for the
evolutionary state of the GH{\sc ii}Rs. The points defining the
vertexes of the triangle should correspond to the three parts of
the shell: center, inner edge and outer edge (see their Fig. 3).
The velocity dispersion in the center reaches the maximum value,
while the minimum value occurs in the outer edge. The inner edge
presents higher intensity and higher velocity dispersion than the
outer edge. Assuming that the pattern drawn in the
Figure~\ref{schem}A represents a young shell, as it ages, $\sigma$
at the center becomes gradually lower, as well as the difference
between the intensity at the center and  at the inner edge. Then,
the whole figure shifts down, and the upper angle becomes smaller.
Furthermore, in their scenario, older shells should present higher
intensities than young ones, and should be mapped at right of the
young ones in the diagram.

Figure~\ref{schem}B shows a vertical band pattern in $I$-$V$
diagram. This pattern indicates the contribution of
resolved systematic radial movements.
Abrupt velocity variations in a short intensity range should
indicate a local expansion of the gas. It is expected that regions
which present the inclined band pattern in the $I$-$\sigma$
diagram also present a well defined vertical band pattern in
$I$-$V$ diagram. In general, any vertical variations ($\Delta V$)
can be used to quantify the component due to expanding motions in
a region. We used the approximation FWHM$_{rad} \sim$
$\Delta V$ to infer the broadening component due to radial
velocity variations ($\sigma_{rad}$) for regions with almost
constant $\sigma$ values and nearly Gaussian emission line
profiles. However, if the emission profiles are not well
represented by single Gaussians, $\Delta V$ measured in $I$-$V$
plot should underestimate the real component due to systematic
radial expansion. As an alternative, one might try to measure the
line splitting from the most asymmetric or double peaked profile,
in order to better estimate $\Delta V$ from two centroids.

Figure~\ref{schem}C shows a correlation between $V$ and $\sigma$.
This pattern indicates systematic motion with a significant component
in the line of sight. The pattern drawn indicates differentiated
group behavior: in this case, gas with relatively high $\sigma$ is
moving away from the observer. Although we have drawn the pattern with a positive
slope, a region with the same systematic motions could be identified
in a negative slope pattern. Such dependence should be expected in
the presence of relative motions between distinct clouds 
of ionized gas with different internal properties.
Champagne flows should reproduce the pattern like the one shown in
Figure~\ref{schem}C (see $\S$ III\textit{c} in \cite{ski84}). The
simplest pattern would be the random distribution of the points in
the plane $V$-$\sigma$, indicating no dependence between the
variables. In this latter case, it may represent isotropic
expansion or turbulence, but with no privileged direction.



\clearpage



\clearpage

\clearpage

\clearpage

\clearpage

\clearpage

\clearpage

\begin{figure*}
\epsscale{0.49} \plotone{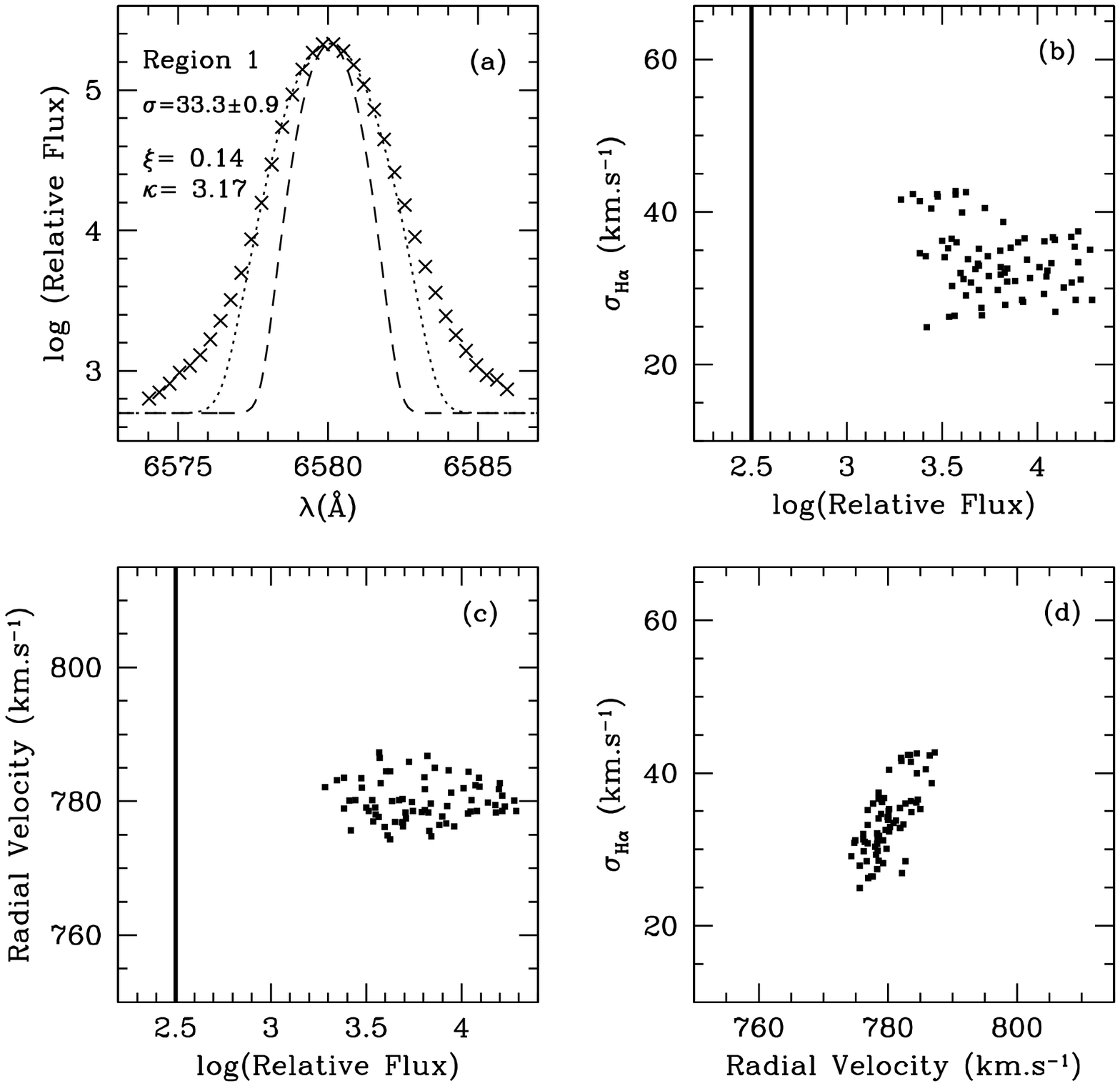}\plotone{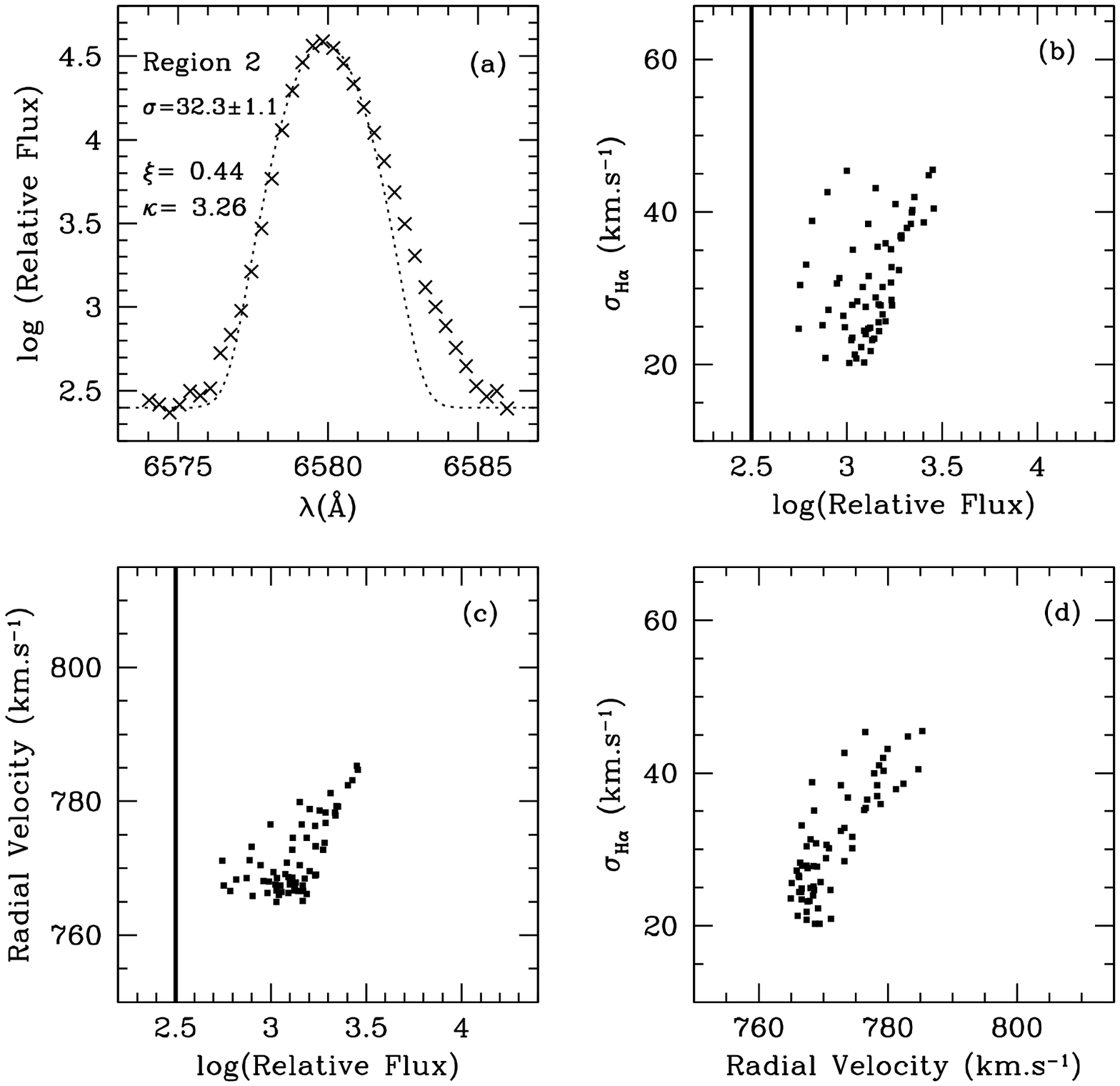}
\plotone{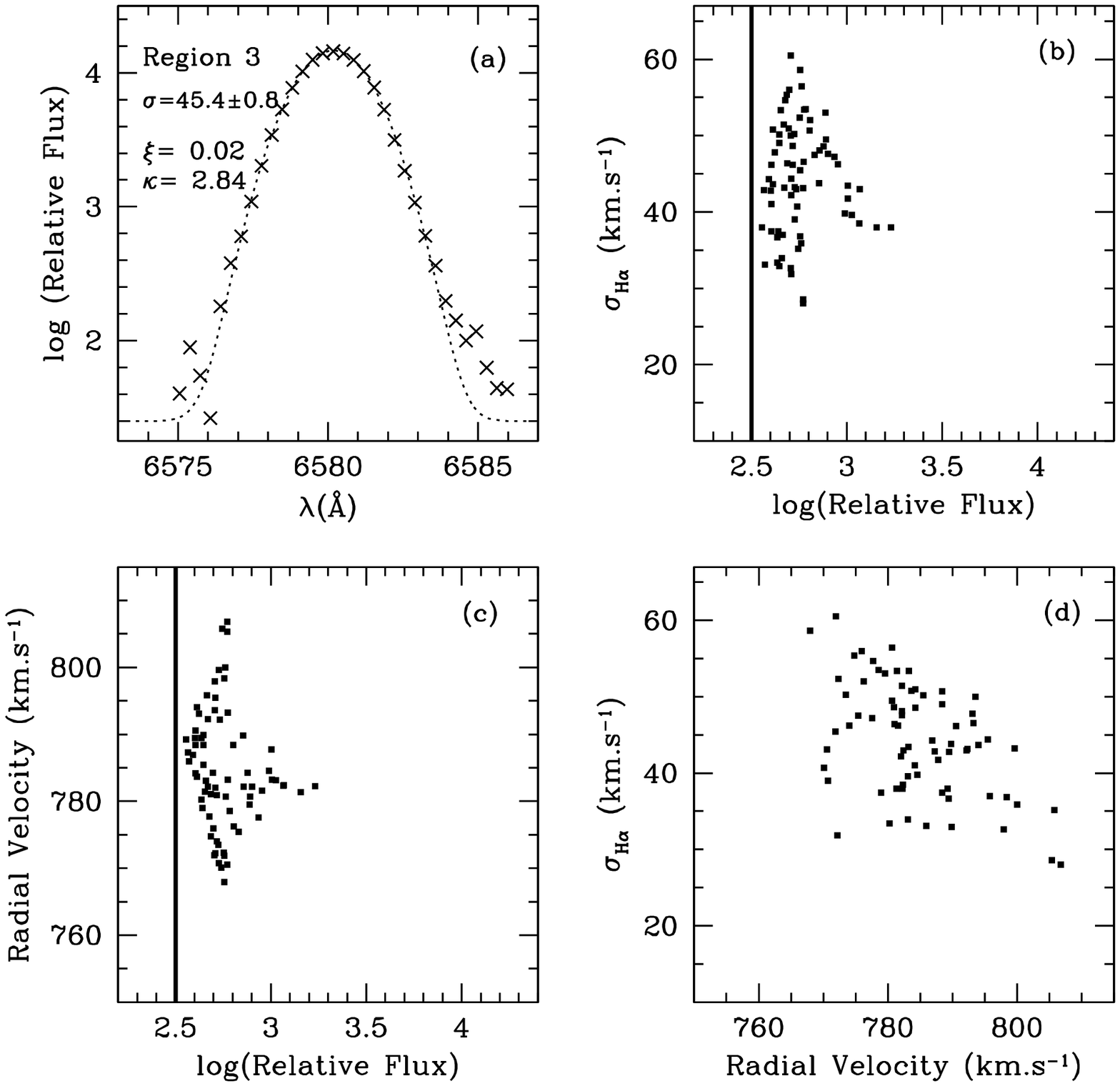}\plotone{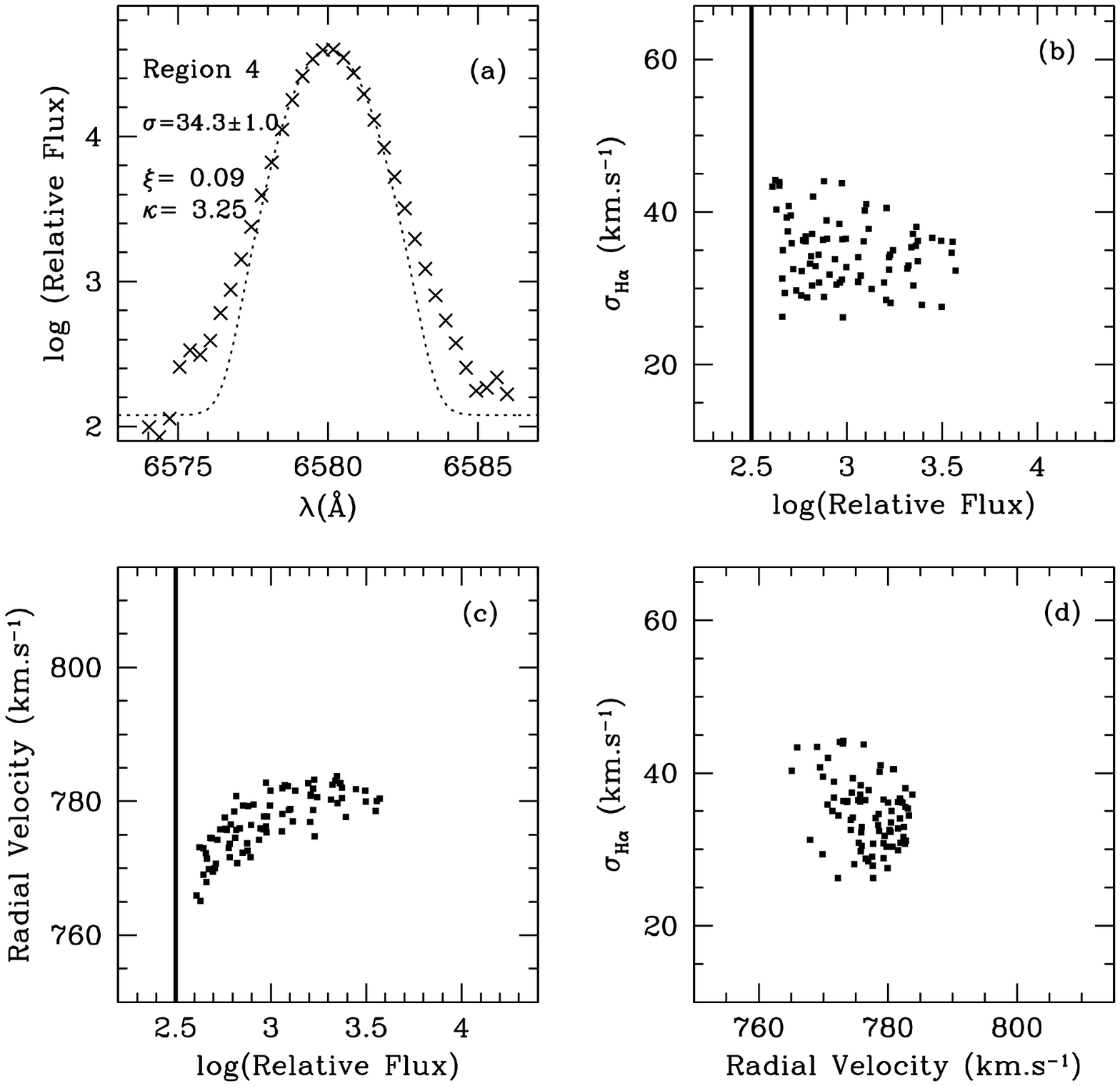} \caption{The panel
present the four boxes discussed in the text for each region. The
dashed line in box (1a) represents the instrumental profile (FWHM
= 1.5{\AA}). The dotted lines in all boxes 1 represent the single
Gaussian fits to the integrated profiles. Note that in
linear-logarithmic axes a Gaussian takes the form of a parabola. The
region number and the $\sigma$ value derived are shown in the boxes
(a). $I$-$\sigma$, $V$-$\sigma$ and $V$-$\sigma$ for each region
are presented by boxes (b), (c) and (d), respectively. The
vertical solid line in boxes (b) and (c) at log(Relative Flux) =
2.5 represent the estimated minimum line flux to well measure a
single line profile.\label{regions}}
\end{figure*}

\clearpage

\addtocounter{figure}{-1}
\begin{figure*}
\epsscale{0.49} \plotone{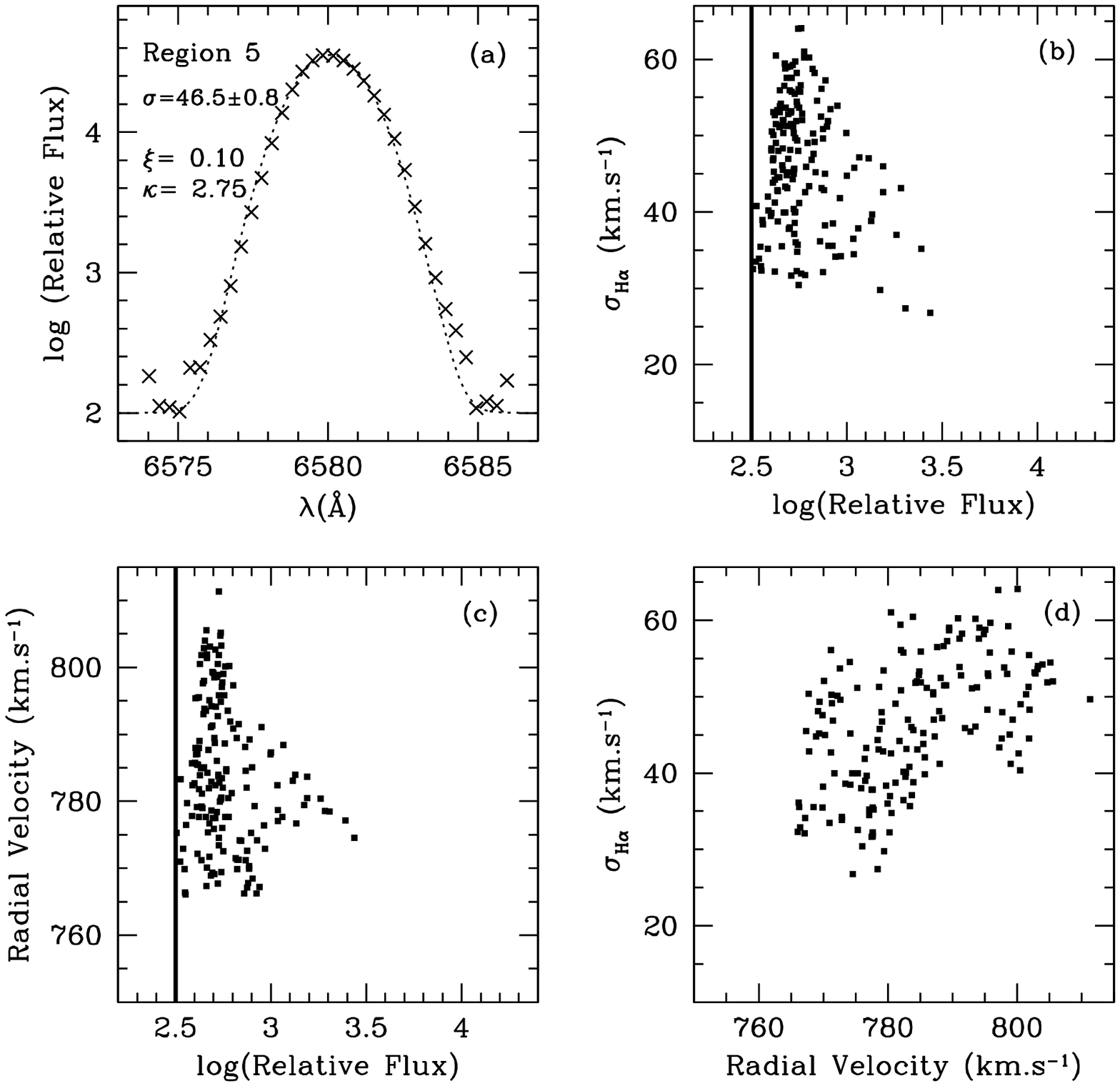}\plotone{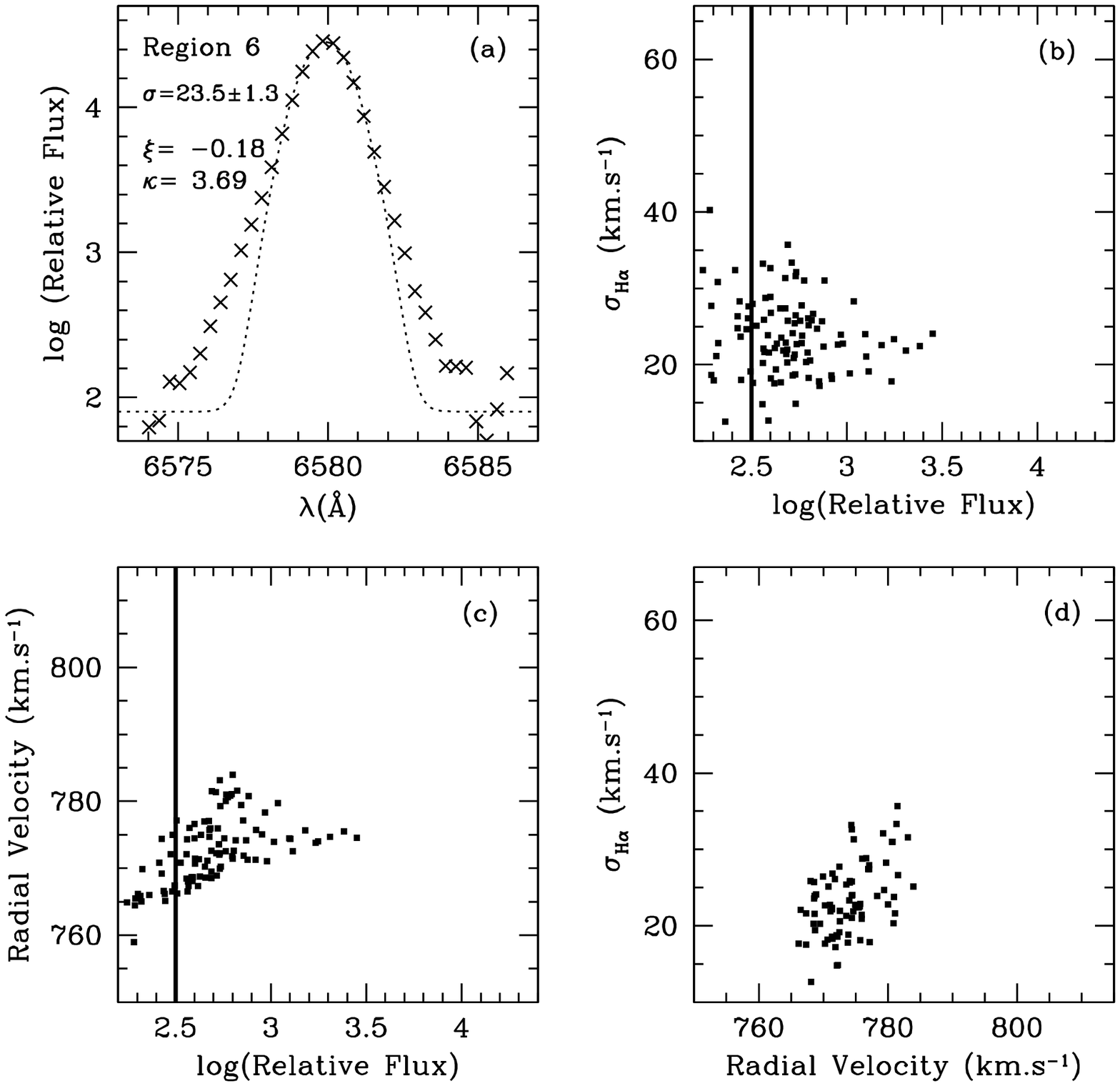}
\plotone{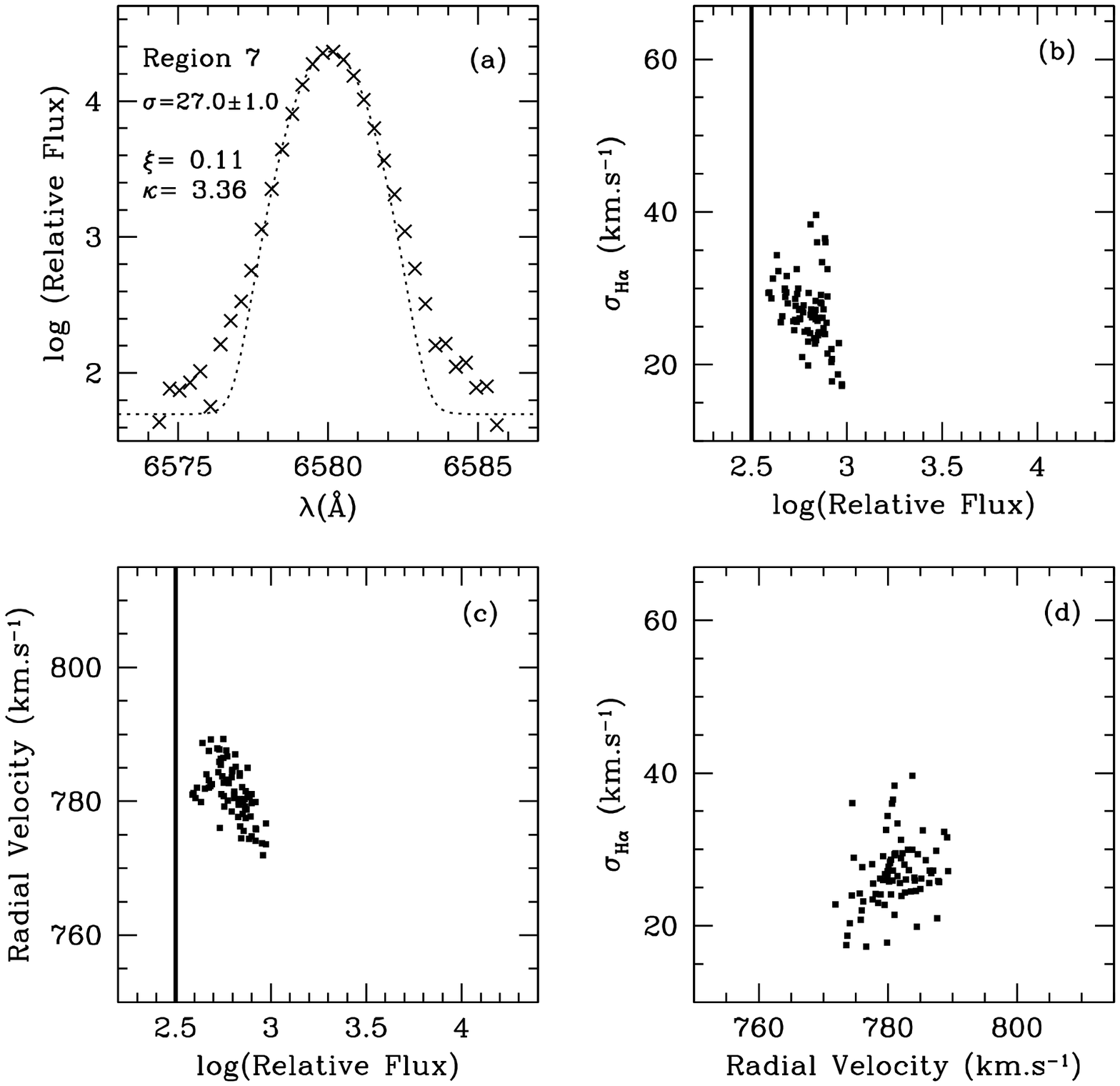}\plotone{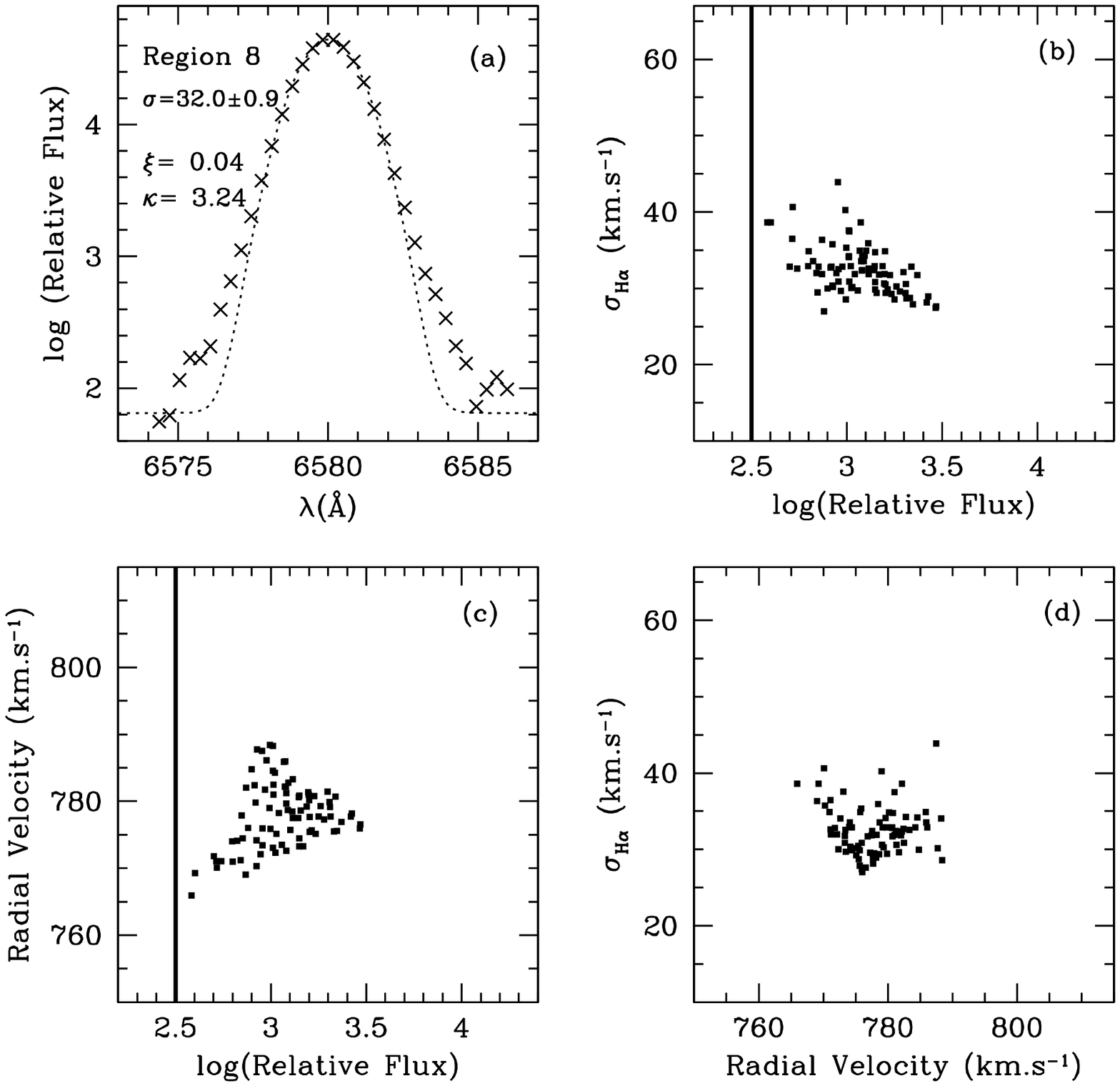} \caption{Continued.
Only Region 6 presentS points with log(Relative Flux) $<$ 2.5 and
they are not shown in box 6b.}
\end{figure*}




\begin{figure}
\epsscale{0.9}\plotone{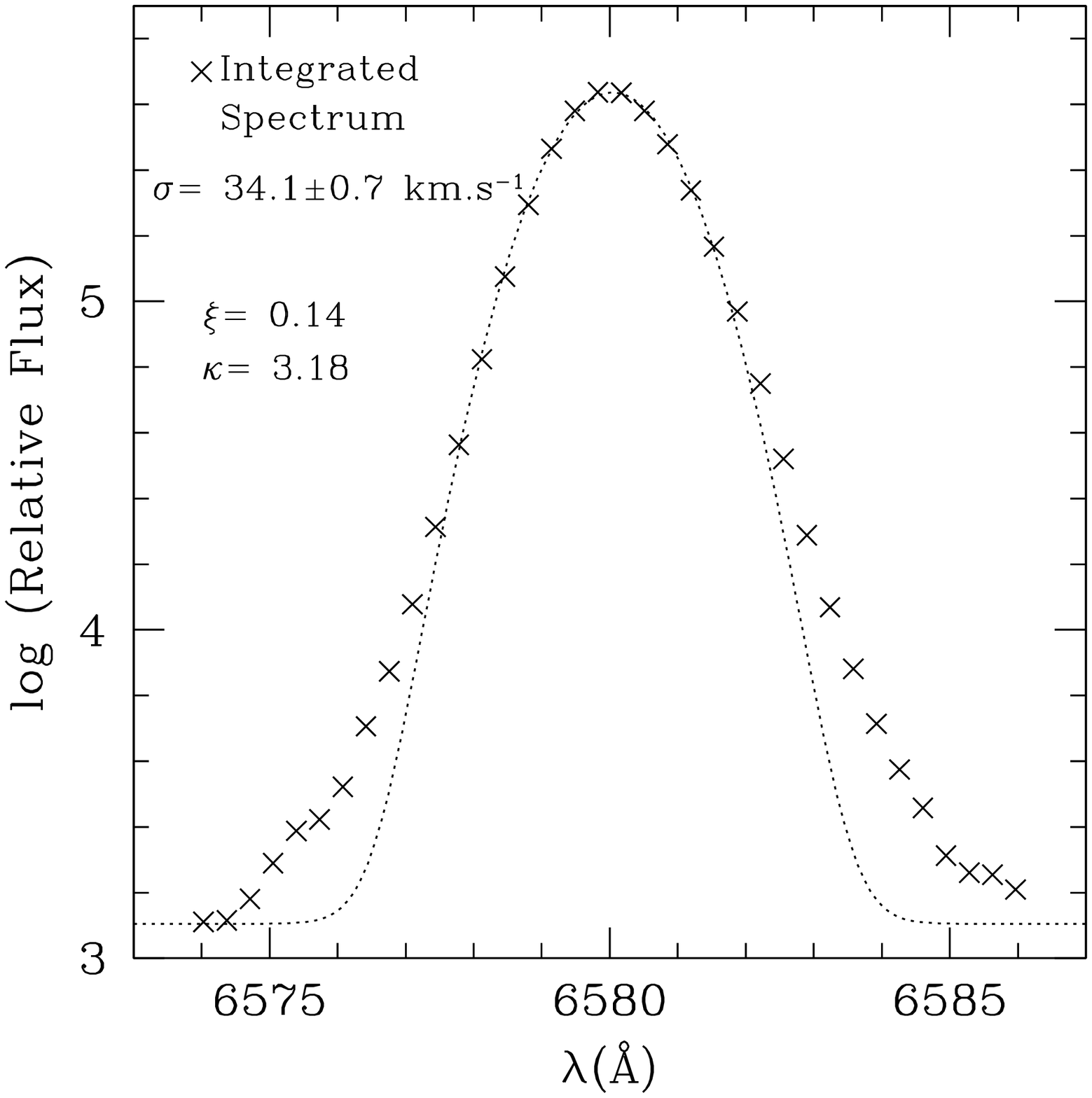}\caption{The fully integrated
profile summing all eight regions analyzed in section~\ref{regs}.
The dotted line represent the single Gaussian fit to the
profile.\label{diag_int}}
\end{figure}

\clearpage

\begin{figure}
\epsscale{0.49} \plotone{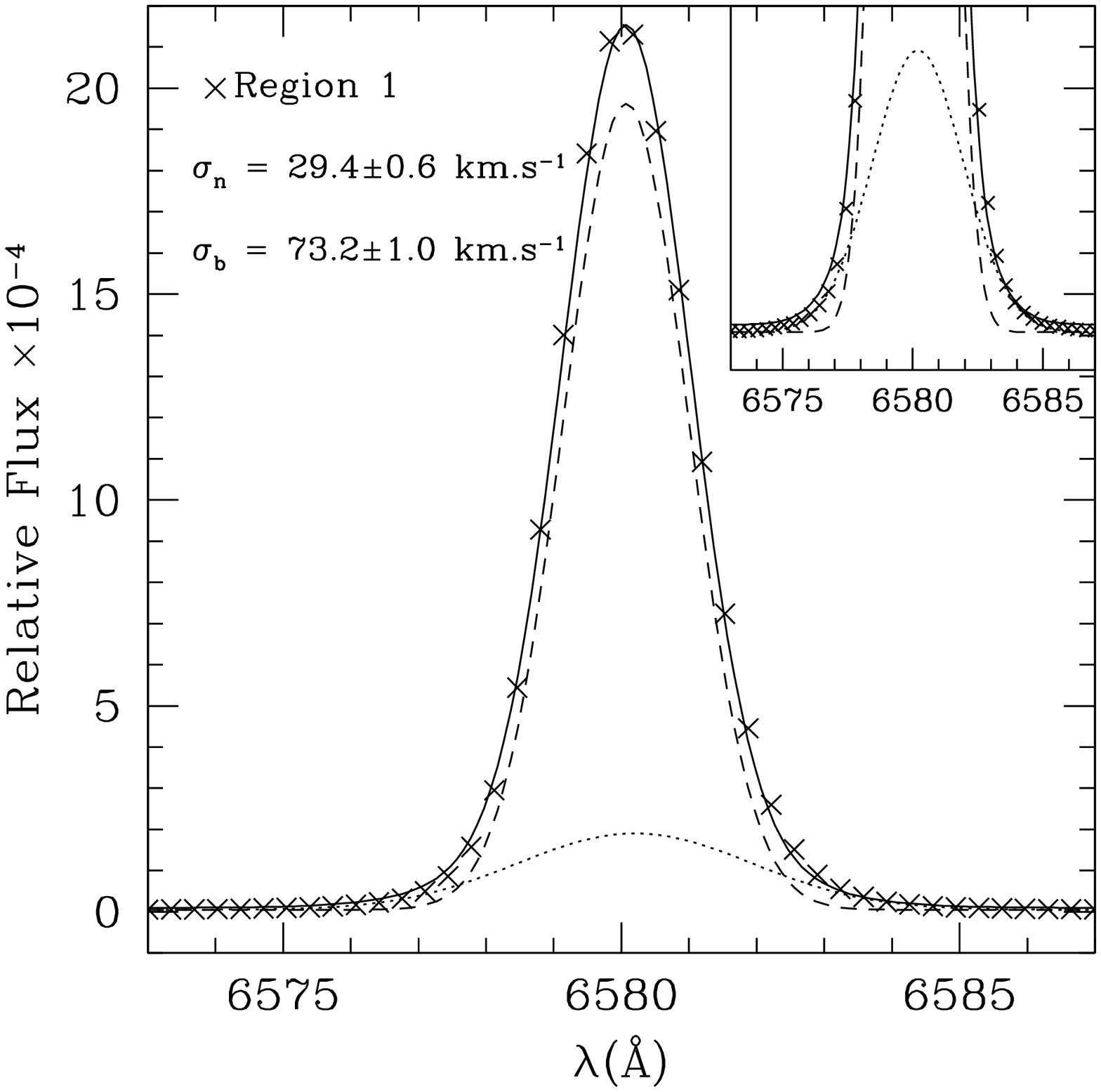} \caption{The integrated
profile of Region 1 (same in Figure~\ref{regions} box 1a) plotted
in linear axes. The profile is well fitted by two Gaussians and
the profiles are plotted. The dashed line represents the narrow
component $\sigma_{n}$, whereas the dotted line represents the
broad component $\sigma_{b}$. The solid line shows the resultant
profile.\label{comp}}
\end{figure}

\clearpage

\begin{figure*}

\epsscale{1}\plotone{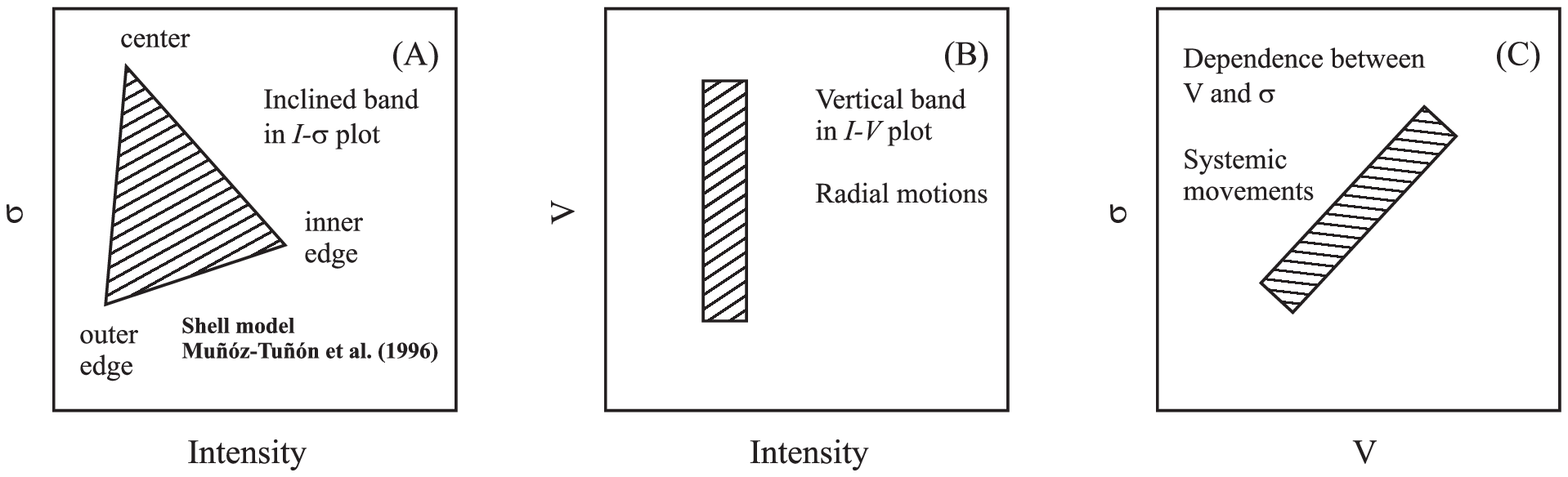}

\caption{Schematic diagrams showing idealized patterns formed by the
distribution of points discussed in the text.\label{schem}}
\end{figure*}


\begin{thebibliography}{}

\bibitem[Baldwin et al.(1982)]{bal82} Baldwin, J. A., Spinrad,
H.,  \& Terlevich, R. 1982, \mnras, 198, 535

\bibitem[Beck et al.(2002)]{bec02} Beck, S.C., Turner, J.L., Langland-Shula,
L.E.,
 Meier, D.S., Crosthwaite, L.P., Gorjian, V. 2002, \aj, 124, 2516


\bibitem[Bosch et al.(2002)]{bos02} Bosch, G., Terlevich, E.,
    \& Terlevich, R. 2002, \mnras, 329, 481

\bibitem[Boulesteix(1999)]{bou99} Boulesteix J. ADHOC User's Manual
http://www-obs.cnrs-mrs.fr/adhoc/adhoc.html

\bibitem[Brinks \& Klein(1988)]{bri88} Brinks, E.,
    \& Klein, U. 1988, \mnras, 231, 63

\bibitem[Castellanos et al.(2002)]{cas02} Castellanos, M., D\'{\i}az, A. I. \&
Terlevich, E.  2002, \mnras, 329, 315

\bibitem[Chu \& Kennicutt(1994)]{chu94} Chu, Y.-H.,
    \& Kennicutt, Jr., R. C. 1994, \apj, 425, 720


\bibitem[Cumming et al.(2008)]{cum08} Cumming, R. J., Fathi, K., Ostlin, G., Marquart, T., M\'arquez, I., Masegosa, J., Bergvall, N., Amram, P., 2008, \aap, 479,725




\bibitem[Dimeo (2002)]{dim05} Dimeo, R. 2005, PAN User Guide,
ftp://ftp.ncnr.nist.gov/pub/staff/dimeo/pandoc.pdf



\bibitem[Denicol\'o et al.(2002)]{den02} Denicol\'o, G., Terlevich, R. \&
Terlevich, E.,  2002, \mnras, 330, 69

\bibitem[Dopita \& Sutherland(2003)]{dop03} Dopita, M.E. \& Sutherland,
R.S.,in
``Astrophysics of the Diffuse Universe'', 2003, Springer-Verlag Berlin
Heidelberg New York

\bibitem[Dyson(1979)]{dys79} Dyson, J. E. 1979, \aap, 73, 132


\bibitem[Fuentes-Masip et al.(2000)]{fue00} Fuentes-Masip, O.,
Mu\~n\'oz-Tu\~n\'on, C., Casta\~neda, H. O., \& Tenorio-Tagle, G.
2000, \aj, 120, 752

\bibitem[Hibbard \& Milhos(1995)]{hib95} Hibbard, J. E.,
    \& Mihos, J. C. 1995, \aj, 110, 140


\bibitem[Kaufer et al.(1999)]{kau99}Kaufer, A., Stahl, O., Tubbesing, S.,
Norregaard, P., Avila, G., Francois, P., Pasquini,L. \& Pizzella,
A. 1999, The Messenger 95, 8

\bibitem[Kehrig et al.(2004)]{keh04} Kehrig, C., Telles, E.,
    \& Cuisinier, F. 2004, \aj, 128, 1141


\bibitem[Kunth \& Sargent(1981)]{kun81} Kunth, D.,
    \& Sargent, W. L. W. 1981, \aap, 101, L5


\bibitem[Ma\'iz-Apell\'aniz et al.(1999)]{mai99} Ma\'iz-Apell\'aniz, J.,
Mu\~noz-Tu\~n\'on, C., Tenorio-Tagle, G., \& Mas-Hesse, J. M. 1999
\aap, 343, 64

\bibitem[Mart\' inez-Delgado et al.(2007)]{mar07} Mart\' inez-Delgado, I.,
Tenorio-Tagle, G., Mu\~noz-Tu\~n\'on, C., Moiseev, A., Cair\'os, L. M., 2007,
\aj, 133, 2892

\bibitem[Melnick(1977)]{mel77} Melnick, J. 1977, \apj, 213, 15

\bibitem[Melnick et al.(1987)]{mel87} Melnick, J., Moles, M., Terlevich, R.,
    \& Garcia-Pelayo, J.-M. 1987, \mnras, 226, 849

\bibitem[Melnick et al.(1988)]{mel88} Melnick, J., Terlevich, R.,
    \& Moles, M. 1988, \mnras, 235, 297

\bibitem[Melnick et al.(1999)]{mel99} Melnick, J.,
Tenorio-Tagle, G.,
    \& Terlevich, R. 1999, \mnras, 302, 677

\bibitem[Miesch \& Scalo(1995)]{ms95} Miesch,M.S. \& Scalo, J.M. 1995, \apjl, 450, L27

\bibitem[M\"unch(1958)]{mun58} M\"unch, G. 1958, Cosmical Gas Dynamics,
Proceedings from IAU Symposium no. 8. Edited by Johannes Martinus Burgers
and Richard Nelson Thomas. International Astronomical Union. Symposium no. 8, p.
1035

\bibitem[Mu\~noz-Tu\~n\'on et al.(1995)]{mun95} Mu\~noz-Tu\~n\'on, C.,
Gavryusev, V., \& Casta\~neda, H. O. 1995, \aj, 110, 1630

\bibitem[Mu\~noz-Tu\~n\'on et al.(1996)]{mun96} Mu\~noz-Tu\~n\'on, C.,
Tenorio-Tagle, G., Casta\~neda, H. O., \& Terlevich, R. 1996, \aj,
112, 1636


\bibitem[\"Ostlin et al.(2004)]{ost04} \"Ostlin, G., Cumming, R. J., Amram, P., Bergvall, N., Kunth, D., Márquez, I., Masegosa, J., Zackrisson, E., 2004, \aap, 419, L43

\bibitem[Pelupessy et al.(2004)]{pel04} Pelupessy, F.I., van der Werf, P.P \&
Icke, V., 2004, \aap, 422, 55




\bibitem[Rozas et al.(2006)]{roz06a} Rozas, M., Richer, M. G.,
L\'opez, J.A., Rela\~no, M., \& Beckman, J. E. 2006, \aap, 455,
539



\bibitem[Roy et al.(1991)]{roy91} Roy, J.-R., Boulesteix, J.,
Joncas, G., \& Grundseth, B. 1991, \apj, 367, 141

\bibitem[Sargent \& Searle(1970)]{sar70} Sargent, W. L. W.
    \& Searle, L. 1970, \apj, 162, 155


\bibitem[Searle \& Sargent(1972)]{sea72} Searle, L.,
    \& Sargent, W. L. 1972, \apj, 173, 25



\bibitem[Skillman \& Balick(1984)]{ski84} Skillman, E. D.,
    \& Balick, B. 1984, \apj, 280, 580

\bibitem[Smith \& Weedman(1970a)]{smi70a} Smith, M. G.,
    \& Weedman, D. W. 1970a, \apj, 161, 33

\bibitem[Smith \& Weedman(1970b)]{smi70b} Smith, M. G.,
    \& Weedman, D. W. 1970b, \apj, 160, 65

\bibitem[Smith \& Weedman(1972)]{smi72} Smith, M. G.,
    \& Weedman, D. W. 1972, \apj, 172, 307

\bibitem[Tacconi \& Young(1987)]{tac87} Tacconi \& Young 1987, ApJ, 322, 681

\bibitem[Telles et al.(2001)]{tel01} Telles, E.,
Mu\~noz-Tu\~n\'on, C., \& Tenorio-Tagle, G. 2001, \apj, 548, 671

\bibitem[Telles \& Terlevich(1995)]{tt95} Telles, E. \& Terlevich, R., 1995, \mnras,
275, 1

\bibitem[Telles \& Terlevich(1997)]{tt97} Telles, E. \& Terlevich, R., 1997, \mnras,
286, 183

\bibitem[Telles, Melnick \& Terlevich(1997)]{tmt97} 
Telles, E., Melnick, J. \& Terlevich R. 1997, \mnras, 288, 78

\bibitem[Telles \& Maddox(2000)]{tm00} Telles, E. \& Maddox, S., 2000, \mnras, 311,
307


\bibitem[Tenorio-Tagle(1979)]{ten79} Tenorio-Tagle, G. 1979, \aap, 71,
59

\bibitem[Tenorio-Tagle et al.(1993)]{ten93} Tenorio-Tagle, G.,
Mu\~noz-Tu\~n\'on, C.,
    \& Cox, D. P. 1993, \apj, 418, 767

\bibitem[Tenorio-Tagle et al.(2006)]{ten06} Tenorio-Tagle, G.,
Mu\~noz-Tu\~n\'on, C.,
P\'rez, E., Silich, S. Telles, E. 2006, \apj, 643, 186

\bibitem[Terlevich et al.(1991)]{ter91} Terlevich, R., Melnick, J., Masegosa,
J.,    \& Moles, M., Copetti, M. V. F. 1991, \aaps, 91, 285

\bibitem[Terlevich \& Melnick(1981)]{ter81} Terlevich, R.,
    \& Melnick, J. 1981, \mnras, 195, 839


\bibitem[Townsley et al.(2006)]{tow06} Townsley, L. K., Broos, P.S., Feigelson,
E.D., Garmire, G.P. \&
 Getman, K.V., 2006, \aj, 131, 2140


\bibitem[Ulvestad et al.(2007)]{ulv07} Ulvestad, J.S., Johnson, K.E. \& Neff, S.G.,
2007, \aj, 133, 1868

\bibitem[Vacca \& Conti(1992)]{vac92} Vacca, W. D.,
    \& Conti, P.S 1992, \apj, 401, 543

\bibitem[van Zee et al.(1998)]{van98} van Zee, L., Skillman, E. D.,
    \& Salzer, J. J. 1998, \aj, 116, 1186

\bibitem[Vanzi et al.(1996)]{van96} Vanzi, L., Rieke, G. H., Martin, C. L.,
    Shields, J. C. 1996, \apj, 466, 150

\bibitem[Vanzi et al.(2008)]{van08} Vanzi, L., Cresci, G., Telles,E. \&
    Melnick, J., 2008, \aap, 486, 393


\bibitem[Walsh \& Roy(1993)]{wal93} Walsh, J. R.,
    \& Roy, J. R. 1993, \mnras, 262, 27

\bibitem[Wang(1999)]{wan99} Wang, Q. D. 1999, \apj, 510, L139


\bibitem[Westmoquette et al.(2007a)]{wes07a} Westmoquette, M.S., Exter, K.M,
Smith, L.J. \&
Gallagher III, J.S., \mnras, 381, 894

\bibitem[Westmoquette et al.(2007b)]{wes07b} Westmoquette, M.S.,  Smith, L.J.,
Gallagher III, J.S. \& Exter, K.M \mnras, 381, 913


\bibitem[Yang et al.(1996)]{yan96} Yang, H., Chu, Y.-H., Skillman, E. D.,
    \& Terlevich, R. 1996, \aj, 112, 146

\bibitem[Yorke et al.(1984)]{yor84} Yorke, H. W., Tenorio-Tagle, G.,
    \& Bodenheimer, P. 1984, \aap, 138, 325


\end{thebibliography}
\end{document}